\documentclass[11pt,english,onecolumn]{IEEEtran}
\usepackage[T1]{fontenc}
\usepackage[latin9]{inputenc}
\usepackage{amsmath}
\usepackage{amsthm}
\usepackage{amssymb}
\usepackage{graphicx}
\usepackage{setspace}
\makeatletter
\theoremstyle{plain}
\newtheorem{thm}{\protect\theoremname}
\theoremstyle{plain}
\newtheorem{prop}[thm]{\protect\propositionname}
\theoremstyle{plain}
\newtheorem{lem}[thm]{\protect\lemmaname}
\theoremstyle{remark}
\newtheorem{rem}[thm]{\protect\remarkname}
\theoremstyle{plain}
\newtheorem{cor}[thm]{\protect\corollaryname}
\theoremstyle{plain}
\newtheorem{conjecture}[thm]{\protect\conjecturename}

\usepackage{hyperref}
\usepackage{breqn}
\usepackage{bbm} 

\allowdisplaybreaks

\DeclareMathOperator{\ord}{ORD}
 \DeclareMathOperator{\pord}{PORD}

\DeclareMathOperator{\maj}{\mathrm{Maj}}
\DeclareMathOperator{\dic}{\mathrm{Dict}}
\DeclareMathOperator{\gdic}{\mathrm{G-Dict}}

\global\long\def\binent{h}

\global\long\def\Hamw{W_H}
\global\long\def\Bin{\mathrm{Bin}}
\global\long\def\lex{\mathrm{lex}}

\global\long\def\s[#1]{\textnormal{\scriptsize #1}}
\global\long\def\st[#1]{\textnormal{\tiny #1}}

\global\long\def\P{\Pr}
\global\long\def\E{\mathbb{E}}

\global\long\def\I{\mathbbm{1}}
\global\long\def\d{\mathrm{d}}
\global\long\def\e{\mathrm{e}}

\global\long\def\dfn{:=}

\global\long\def\trre[#1,#2]{\overset{{\scriptstyle (#2)}}{#1}} 

\author{Nir Weinberger, \IEEEmembership{Member, IEEE,} and Ofer Shayevitz, \IEEEmembership{Senior Member, IEEE}}

\makeatother

\usepackage{babel}
\providecommand{\conjecturename}{Conjecture}
\providecommand{\corollaryname}{Corollary}
\providecommand{\lemmaname}{Lemma}
\providecommand{\propositionname}{Proposition}
\providecommand{\remarkname}{Remark}
\providecommand{\theoremname}{Theorem}

\begin{document}

\title{Guessing with a Bit of Help\thanks{The authors are with the Department of EE--Systems, Tel Aviv University, Tel Aviv, Israel. Emails: \{nir.wein@gmail.com, ofersha@eng.tau.ac.il\}. This work was supported by an ERC grant no. 639573.}}

\maketitle
\renewcommand\[{\begin{equation}}
\renewcommand\]{\end{equation}}
\begin{abstract}
What is the value of a single bit to a guesser? We study this problem
in a setup where Alice wishes to guess an i.i.d. random vector, and
can procure one bit of information from Bob, who observes this vector
through a memoryless channel. We are interested in the \emph{guessing
efficiency}, which we define as the best possible multiplicative reduction
in Alice's guessing-moments obtainable by observing Bob's bit. For
the case of a uniform binary vector observed through a binary symmetric
channel, we provide two lower bounds on the guessing efficiency by
analyzing the performance of the Dictator and Majority functions,
and two upper bounds via maximum entropy and Fourier-analytic / hypercontractivity
arguments. We then extend our maximum entropy argument to give a lower
bound on the guessing efficiency for a general channel with a binary
uniform input, via the strong data-processing inequality constant
of the reverse channel. We compute this bound for the binary erasure
channel, and conjecture that Greedy Dictator functions achieve the
guessing efficiency.
\end{abstract}

\section{Introduction}

In the classical problem of guessing, Alice wishes to learn the value
of a discrete random variable (r.v.) $X$ as quickly as possible,
by sequentially asking yes/no questions of the form ``is $X=x$?'',
until she gets it right. Alice's guessing strategy, which is the ordering
of the alphabet of $X$ according to which she states her guesses,
induces a random guessing time. It is well known and simple to check
that the optimal guessing strategy, which simultaneously minimizes
all the positive moments of the guessing time, is to guess according
to decreasing order of probability. Formally then, for any $s>0$,
the \emph{minimal $s$th-order guessing-time moment} of $X$ is 
\begin{equation}
G_{s}(X)\dfn\E\left[\ord_{X}^{s}(x)\right],\label{eq: guessing moment}
\end{equation}
where $\ord_{X}(x)$ returns the index of the symbol $x$ relative
to a the order induced by sorting the probabilities in a descending
order, with ties broken arbitrarily. For brevity, we refer to $G_{s}(X)$
as the \emph{guessing-moment} of $X$.

The guessing problem was first introduced and studied in an information-theoretic
framework by Massey \cite{massey1994guessing}, who drew a relation
between the average guessing time of an r.v. and its entropy, and
was later explored more systematically by Arikan \cite{arikan1996inequality}.
Several motivating problems for studying guesswork are fairness in
betting games, computational complexity of lossy source coding and
database search algorithms (see the introduction of \cite{arikan1998guessing}
for a discussion), secrecy systems \cite{merhav1999shannon,hayashi2008coding,hanawal2011shannon},
crypt-analysis (password cracking) \cite{yona2016password,christiansen2015multi},
and computational complexity of sequential decoding \cite{arikan1996inequality}.
In \cite{arikan1996inequality}, Arikan introduced the problem of
guessing with side information, where Alice is in possession of another
r.v. $Y$ that is jointly distributed with $X$. In that case, the
optimal conditional guessing strategy is to guess by decreasing order
of conditional probabilities, and hence the associated \emph{minimal
conditional $s$th-order guessing-time moment of $X$ given $Y$}
is 
\[
G_{s}(X|Y)\dfn\E\left(\ord_{X|Y}^{s}(X\mid Y)\right),
\]
where $\ord_{X|Y}(x\mid y)$ returns the index of $x$ relative to
a the order induced by sorting the conditional probabilities of $X$,
given that $Y=y$, in a descending order. Arikan showed that, as intuition
suggests, side information reduces the guessing-moments \cite[Corollary 1]{arikan1996inequality}
\[
G_{s}(X|Y)\leq G_{s}(X).
\]
Furthermore, he showed that if $\{(X_{i},Y_{i})\}_{i=1}^{n}$ is an
i.i.d. sequence, then \cite[Prop. 5]{arikan1996inequality} 
\[
\lim_{n\to\infty}\frac{1}{n}\log G_{s}^{1/s}(X^{n}|Y^{n})=H_{\frac{1}{1+s}}(X_{1}\mid Y_{1}),
\]
where $H_{\alpha}(X\mid Y)$ is the Arimoto-Rényi conditional entropy
of order $\alpha$. The information-theoretic analysis of the guessing
problem was further extended in multiple directions, such as allowing
distortion in the guess \cite{arikan1998guessing}, guessing under
source uncertainty \cite{sundaresan2007guessing}, improved bounds
at finite blocklength \cite{boztas1997comments} and an information-spectrum
analysis \cite{hanawal2011guessing}, to name a few.

In the conditional setting described above, one may think of $Y^{n}$
as side information observed by a \textquotedbl{}helper\textquotedbl{},
say Bob, who then sends his observations to Alice. In this case, as
in the problem of source coding with a helper \cite{wyner1973theorem,ahlswede1975source},
it is more realistic to impose some communication constraints, and
assume that Bob can only send a compressed description of $Y^{n}$
to Alice. This question was recently addressed by Graczyk and Lapidoth
\cite{Graczyk_Lapidoth,gr:17:thesis}, who considered the case where
Bob encodes $Y^{n}$ at a positive rate using $nR$ bits, before sending
this description to Alice. They then characterized the best possible
guessing-moments attained by Alice for general distributions, as a
function of the rate $R$. In this paper, we take this setting to
its extreme, and attempt to quantify the value of a\emph{ single bit}
in terms of reducing the guessing-moments, by allowing Bob to use
only a one-bit description of $Y^{n}$. To that end, we define (in
Section \ref{sec:Problem-Statement}) and study the \emph{guessing
efficiency}, which is the (asymptotically) best possible multiplicative
reduction in the guessing-moments of $X^{n}$ offered by observing
a Boolean function $f(Y^{n})$, i.e., the minimal possible ratio $G_{s}(X^{n}\mid f(Y^{n}))/G_{s}(X^{n})$
as a function of $s$, in the limit of large $n$.

Characterizing the guessing efficiency appears to be a difficult problem
in general. Here we mostly focus on the special case where $X^{n}$
is uniformly distributed over the Boolean cube $\{0,1\}^{n}$, and
$Y^{n}$ is obtained by passing $X^{n}$ through a memoryless binary
symmetric channel (BSC) with crossover probability $\delta$. We derive
two upper bounds and two lower bounds on the guessing efficiency in
this case. The upper bounds, presented in Section \ref{sec:Achievable-Bounds},
are derived by analyzing the efficiency attained by two specific functions,
\emph{Dictator} and \emph{Majority}. We show that neither of these
functions is better than the other for all values of the moment order
$s$. The two lower bounds, presented in Section \ref{sec:Converse-Bounds},
are based on relating the guessing-moment to entropy using maximum-entropy
arguments (generalizing \cite{massey1994guessing}), and on Fourier-analytic
techniques together with a hypercontractivity argument \cite{Bool_book}.
Several graphs illustrating the bounds are given in Section \ref{sec: Graphs}.
In Section \ref{sec: General source-channel} we briefly discuss the
more general case where $X^{n}$ is still uniform over the Boolean
cube, but $Y^{n}$ is obtained from $X^{n}$ via a general binary-input,
arbitrary-output channel. We generalize our entropy lower bound to
this case using the strong data-processing inequality (SDPI) applied
to the reverse channel (from $Y$ to $X$). We then discuss the case
of the binary erasure channel (BEC), for which we also provide an
upper bound by analyzing the \emph{Greedy Dictator} function, namely
where Bob sends the first bit that has not been erased. We conjecture
that this function minimizes the guessing efficiency simultaneously
at all erasure parameters and all moments $s$.

\textit{Related Work.} Graczyk and Lapidoth \cite{Graczyk_Lapidoth,gr:17:thesis}
considered the same guessing question in the case where Bob can communicate
with Alice at some positive rate $R$, i.e., can use $nR$ bits to
describe $Y^{n}$. This setup facilitates the use of large-deviation-based
information-theoretic techniques, which allowed the authors to characterized
the optimal reduction in the guessing-moments as a function of $R$.
We note that this type of random-coding arguments cannot be applied
in our extermal one-bit setup. Characterizing the guessing efficiency
in the case of the BSC with a uniform input can also be thought of
as a guessing variant of the \emph{most informative Boolean function
problem} introduced by Kumar and Courtade \cite{Boolean_conjecture},
who have asked about the maximal reduction in the entropy of $X^{n}$
obtainable by observing a Boolean function $f(Y^{n})$. They have
conjectured that a \emph{Dictator} \emph{function, }e.g. $f(y^{n})=y_{1}$
is optimal simultaneously at all noise levels, see \cite{Ordentlich_Shayevitz_Weinstein,Alex,kindler2015remarks,li2018boolean}
for some recent progress. We note that as in the guessing case, allowing
Bob to describe $Y^{n}$ using $nR$ bits renders the problem amenable
to an exact information-theoretic characterization \cite{Chandar_Tcham}.
In another related work \cite{Boolean_quadratic}, we have asked about
the Boolean function $Y^{n}$ that maximizes the reduction in the
sequential mean-squared prediction error of $X^{n}$, and have shown
that the Majority function is optimal in the noiseless case, yet that
there is no single function that is simultaneously optimal at all
noise levels. Finally, in a recent work \cite{burin2017reducing}
the average guessing time using the help of a noisy version of $f(X^{n})$,
has been considered. By contrast, in this paper the noise is applied
to the inputs of the function, rather than to its output.

\section{Problem Statement\label{sec:Problem-Statement}}

Let $X^{n}$ be an i.i.d. vector from distribution $P_{X}$, who is
transmitted over a memoryless channel of conditional distribution
$P_{Y|X}$. Bob observes $Y^{n}$ at the output of the channel, and
can send one bit $f:\{0,1\}^{n}\to\{0,1\}$ to Alice, who in turn
needs to guess $X^{n}$. Our goal is to characterize the best possible
multiplicative reduction in guessing-moments offered by a function
$f$, in the limit of large $n$. Precisely, we wish to characterize
the \emph{guessing efficiency}, defined as 
\begin{equation}
\gamma_{s}(P_{X},P_{Y|X})\dfn\limsup_{n\to\infty}\min_{f:\{0,1\}^{n}\to\{0,1\}}\frac{G_{s}(X^{n}\mid f(Y^{n}))}{G_{s}(X^{n})}.\label{eq: guessing efficiency optimal}
\end{equation}
In this paper we are mostly interested in the case where $P_{X}=(1/2,1/2)$,
i.e., $X^{n}$ is uniformly distributed over $\{0,1\}^{n}$, and where
the channel is a BSC with crossover probability $\delta\in[0,1/2]$.
With a slight abuse of notation, we denote the guessing efficiency
in this case by $\gamma_{s}(\delta)$. Before we proceed, we note
the following simple facts. 
\begin{prop}
\label{prop: general properties}The following claims hold: 
\begin{enumerate}
\item For $\gamma_{s}(\delta)$ the limit-supremum (\ref{eq: guessing efficiency optimal})
is a regular limit, achieved by a sequence of deterministic functions. 
\item $\gamma_{s}(\delta)$ is a non-decreasing function of $\delta\in[0,1/2]$
satisfying $\gamma_{s}(0)=2^{-s}$ and $\gamma_{s}(1/2)=1$, where
$\gamma_{s}(0)$ is attained by any sequence of balanced functions. 
\end{enumerate}
\end{prop}
\begin{IEEEproof}
See Appendix \ref{sec: Proofs}. 
\end{IEEEproof}

\section{Upper Bounds on $\gamma_{s}(\delta)$ \label{sec:Achievable-Bounds}}

In this section we derive two upper bounds on the BSC guessing efficiency
$\gamma_{s}(\delta)$, by analyzing two simple functions - the \emph{Dictator}
function and the \emph{Majority} function. Let $a,b\in\mathbb{N}$,
$a\leq b$ be given. The following sum will be useful for the derivations
in the rest of the paper:
\[
K_{s}(a,b)\dfn\frac{1}{b-a}\sum_{i=a+1}^{b}i^{s},
\]
where we will abbreviate $K_{s}(b)\dfn K_{s}(0,b)$. For a pair of
sequences $\{a_{n}\}_{n=1}^{\infty}$,\{$b_{n}\}_{n=1}^{\infty}$
we will Let $a_{n}\cong b_{n}$ mean that $\lim_{n\to\infty}\frac{a_{n}}{b_{n}}=1$. 
\begin{lem}
\label{lem: moments of ordinal}Let $\{a_{n}\}_{n=1}^{\infty}$,\{$b_{n}\}_{n=1}^{\infty}$
be non-decreasing integer sequences such that $a_{n}<b_{n}$ for all
$n$ and $\lim_{n\to\infty}(a_{n}+1)/b_{n}=0$. Then, 
\[
K_{s}(a_{n},b_{n})\cong\frac{1}{s+1}\cdot\frac{\left[b_{n}^{s+1}-a_{n}^{s+1}\right]}{b_{n}-a_{n}}.
\]
Specifically, $G_{s}(X^{n})=K_{s}(2^{n})\cong\frac{2^{sn}}{s+1}$.
\end{lem}
\begin{IEEEproof}
See Appendix \ref{sec: Proofs}.
\end{IEEEproof}
\begin{thm}
\label{thm: dictator}We have
\[
\gamma_{s}(\delta)\leq(1-2\delta)\cdot2^{-s}+2\delta,
\]
 and guessing efficiency equal to the right-hand side (r.h.s.) can
be achieved by a Dictator function.
\end{thm}
\begin{IEEEproof}
Assume without loss of generality that $f(y^{n})=y_{1}$. As $0<\delta<1/2$
it is easily verified that given $y_{1}$, the optimal guessing strategy
is to first guess one of the $2^{n-1}$ vectors for which $x_{1}=y_{1}$
(in an arbitrary order), and then guess one of the remaining $2^{n-1}$
vectors (again, in an arbitrary order). From symmetry, and Lemma \ref{lem: moments of ordinal}
\begin{align}
G_{s}(X^{n}\mid\dic(Y^{n})) & =G_{s}(X^{n}\mid Y_{1}=1)\\
 & =(1-\delta)\cdot K_{s}(2^{n-1})+\delta\cdot K_{s}(2^{n-1},2^{n})\\
 & \cong(1-\delta)\cdot\frac{2^{s(n-1)}}{s+1}+\delta\cdot\frac{2^{s(n-1)}}{s+1}\cdot\left(2^{s+1}-1\right)\\
 & =\frac{2^{s(n-1)}}{s+1}\cdot\left(1-2\delta+\delta\cdot2^{s+1}\right).
\end{align}
The result then follows from (\ref{eq: guessing efficiency optimal})
and Lemma \ref{lem: moments of ordinal}.
\end{IEEEproof}
We next consider the guessing efficiency of the Majority function.
\begin{thm}
\label{thm: majority}Let $\beta\dfn\frac{1-2\delta}{\sqrt{4\delta(1-\delta)}}$
and $Z\sim{\cal N}(0,1)$. Then, 
\begin{equation}
\gamma_{s}(\delta)\leq2\cdot(s+1)\cdot\E\left[Q(\beta Z)\cdot\left(1-Q(Z)\right)^{s}\right],\label{eq: majority guessin efficiency}
\end{equation}
where $Q(\cdot)$ is the tail distribution function of the standard
normal distribution, and guessing efficiency equal to the r.h.s. of
(\ref{eq: majority guessin efficiency}) can be achieved by the Majority
function.
\end{thm}
\begin{IEEEproof}
We assume for simplicity that $n$ is odd. The analysis for an even
$n$ is not fundamentally different. In this case, $f(y^{n})=\maj(y^{n})=\I(\sum_{i=1}^{n}y_{i}>n/2)$,
where $\I(\cdot)$ is the indicator function. To evaluate the guessing-moment,
we first need to find the optimal guessing strategy. To this end,
we let $\Hamw(x^{n})$ be the Hamming weight of $x^{n}$ and note
that the posterior probability is given by
\begin{align}
\P(X^{n}=x^{n}\mid\maj(Y^{n})=1) & =\frac{\P(\maj(Y^{n})=1\mid X^{n}=x^{n})\cdot\P(X^{n}=x^{n})}{\P(\maj(Y^{n})=1)}\\
 & =2^{1-n}\cdot\P\left(\sum_{i=1}^{n}Y_{i}>n/2\mid X^{n}=x^{n}\right)\\
 & =2^{1-n}\cdot\P\left(\sum_{i=1}^{n}Y_{i}>n/2\mid\Hamw(X^{n})=\Hamw(x^{n})\right)\label{eq: posterior majority symmetry}\\
 & \dfn2^{1-n}\cdot r_{n}(\Hamw(x^{n})),
\end{align}
where (\ref{eq: posterior majority symmetry}) follows from symmetry.
Evidently, $r_{n}(w)$ is an increasing function of $w\in\{0,1,\ldots,n\}$.
Indeed, $\Bin(n,\delta)$ be a binomial r.v. of $n$ trials and success
probability $\delta$. Then, for any $w\leq n-1$, as $\delta\leq1/2$,
\begin{align}
 & r_{n}(w+1)\nonumber \\
 & =\P\left(\Bin(w+1,1-\delta)+\Bin(n-w-1,\delta)>n/2\right)\\
 & =\P\left(\Bin(w,1-\delta)+\Bin(1,1-\delta)+\Bin(n-w-1,\delta)>n/2\right)\\
 & \geq\P\left(\Bin(w,1-\delta)+\Bin(1,\delta)+\Bin(n-w-1,\delta)>n/2\right)\\
 & =\P\left(\Bin(w,1-\delta)+\Bin(n-w,\delta)>n/2\right)\\
 & =r_{n}(w),
\end{align}
where in each of the above probabilities, the summations is over independent
binomial r.v's. Hence, we deduce that whenever $\maj(Y^{n})=1$ (resp.
$\maj(Y^{n})=0$) the optimal guessing strategy is by decreasing (resp.
increasing) Hamming weight (with arbitrary order for inputs of equal
Hamming weight). 

We can now turn to evaluate the guessing-moment for the optimal strategy
given Majority. Let $M_{-1}=0$ and $M_{w}=M_{w-1}+\binom{n}{w}$
for $w\in\{0,1,\ldots,n\}$. From symmetry,
\begin{align}
G_{s}(X^{n}\mid\maj(Y^{n})) & =G_{s}(X^{n}\mid\maj(Y^{n})=1)\\
 & =\sum_{w=0}^{n}\binom{n}{w}2^{1-n}r_{n}(w)\sum_{i=M_{w-1}+1}^{M_{w}}i^{s}.
\end{align}
Thus,
\begin{align}
G_{s}(X^{n}\mid\maj(Y^{n})) & \geq\sum_{w=0}^{n}\binom{n}{w}2^{1-n}r_{n}(w)M_{w-1}^{s}\label{eq: majority guessing lower bound}\\
 & =2^{sn+1}\cdot\E\left[r_{n}(W)\left(\frac{M_{W-1}}{2^{n}}\right)^{s}\right]\\
 & =2^{sn+1}\cdot\E\left[r_{n}(W)\P\left(W'\leq W-1\right)^{s}\right],
\end{align}
where $W,W'\sim\Bin(n,1/2)$ and independent. We next evaluate this
expression using the central-limit theorem. To evaluate this expression
asymptotically, we note that the Berry-Esseen theorem \cite[Chapter XVI.5, Theorem 2]{Feller}
leads to (see, e.g., \cite[proof of Lemma 15]{Boolean_quadratic})
\[
r_{n}(w)=Q\left(\beta\cdot\frac{2}{\sqrt{n}}\left[\frac{n}{2}-w\right]\right)+\frac{a_{\delta}}{\sqrt{n}},
\]
for some universal constant $a_{\delta}$. Using the Berry-Esseen
central-limit theorem again, we have that $\frac{2}{\sqrt{n}}(\frac{n}{2}-W')\xrightarrow{d}Z$,
where $Z\sim{\cal N}(0,1)$. Thus for a given $w$,
\begin{align}
\P\left(W'\leq w-1\right) & =1-\P\left(\frac{2}{\sqrt{n}}\left(\frac{n}{2}-W'\right)\geq\frac{2}{\sqrt{n}}\left(\frac{n}{2}-w-1\right)\right)\\
 & =1-Q\left(\frac{2}{\sqrt{n}}\left(\frac{n}{2}-w-1\right)\right)-\frac{a_{1/2}}{\sqrt{n}}\\
 & =1-Q\left(\frac{2}{\sqrt{n}}\left(\frac{n}{2}-w\right)\right)-O\left(\frac{1}{\sqrt{n}}\right),
\end{align}
where the last equality follows from the fact that $|Q'(t)|\leq\frac{1}{\sqrt{2\pi}}$.
Using the Berry-Esseen theorem once again, we have that $\frac{2}{\sqrt{n}}(\frac{n}{2}-w)\xrightarrow{d}Z$,
where $Z\sim{\cal N}(0,1)$. Hence, Portmanteau\textquoteright s lemma
(e.g. \cite[Chapter VIII.1, Theorem 1]{Feller}), and the fact the
$Q(t)$ is continuous and bounded results in
\begin{equation}
G_{s}(X^{n}\mid\maj(Y^{n}))\geq2^{sn+1}\cdot\E\left[Q\left(\beta N\right)\cdot\left(1-Q(N)\right)^{s}\right]+O\left(\frac{1}{n^{s/2}}\right).\label{eq: majority lower bound asymptotic}
\end{equation}
Similarly to (\ref{eq: majority guessing lower bound}), the upper
bound
\[
G_{s}(X^{n}\mid\maj(Y^{n}))\leq\sum_{w=0}^{n}\binom{n}{w}2^{1-n}r_{n}(w)M_{w}^{s},
\]
holds, and a similar analysis leads to an expression which asymptotically
coincides with the r.h.s. of (\ref{eq: majority lower bound asymptotic}).
The result then follows from (\ref{eq: guessing efficiency optimal})
and Lemma \ref{lem: moments of ordinal}.
\end{IEEEproof}
We remark that the guessing efficiency of functions similar to Dictator
and Majority, such as Dictator on $k>1$ inputs ($f(y^{n})=1$ if
and only if $(y_{1},\ldots y_{k})=1^{k}$), or unbalanced Majority
($f(y^{n})=\I(\sum_{i=1}^{n}y_{i}>t)$ for some $t$) may also be
analyzed in a similar way. However, numerical computations indicate
that they do not improve the bounds obtained in Theorems \ref{thm: dictator}
and \ref{thm: majority}, and thus their analysis is omitted.

\section{Lower Bounds on $\gamma_{s}(\delta)$\label{sec:Converse-Bounds}}

We derive two lower bounds on the BSC guessing efficiency $\gamma_{s}(\delta)$,
one based on maximum-entropy arguments, and the other based on Fourier-analytic
arguments.

\subsection{A Maximum-Entropy Bound}
\begin{thm}
\label{thm: maximum entropy}We have
\[
\gamma_{s}(\delta)\geq e^{-1}\cdot\frac{s^{s-1}\cdot(s+1)}{\Gamma^{s}(\frac{1}{s})}\cdot2^{-s(1-2\delta)^{2}}.
\]
\end{thm}
\begin{IEEEproof}
With a standard abuse of notation, let us write the guessing-moment
and the entropy as a function of the distribution. Consider the following
maximum entropy problem \cite[Ch. 12]{Cover:2006:EIT:1146355}
\[
\max_{P\colon G_{s}(P)=g}H(P),
\]
where it should be noted that the support of $P$ is only restricted
to be countable. Assuming momentarily that the entropy is measured
in nats, it is easily verified (using the theory of exponential families
\cite[Ch. 3]{wainwright2008graphical} or by standard Lagrangian duality
\cite[Ch. 5]{Boyd}), that the entropy maximizing distribution is
\[
P_{\lambda}(i)\dfn\frac{\exp(-\lambda i^{s})}{Z(\lambda)}
\]
for $i\in\mathbb{N}_{+}$ where $Z(\lambda)\dfn\sum_{i=1}^{\infty}\exp(-\lambda i^{s})$
is the \emph{partition function}, and $\lambda>0$ is chosen such
that $G_{s}(P_{\lambda})=g$. Thus, the resulting maximum entropy
is given in a parametric form as
\begin{equation}
H(P_{\lambda})=\lambda G_{s}(P_{\lambda})+\ln Z(\lambda).\label{eq: maximum entropy parametric}
\end{equation}
Evidently, if $g=G_{s}(P_{\lambda})\to\infty$ then $\lambda\to0$.
In this case, we may approximate the partition function for $\lambda\to0$
by a Riemann integral. Specifically, by the monotonicity of $e^{-\lambda i^{s}}$
in $i\in\mathbb{N}$, 

\begin{align}
Z(\lambda) & =\sum_{i=1}^{\infty}e^{-\lambda i^{s}}\\
 & =\frac{1}{2}\left(\sum_{i=-\infty}^{\infty}\exp\left(-\left(\frac{|i|}{\lambda^{-1/s}}\right)^{s}\right)-1\right)\\
 & \geq\frac{1}{2}\left(\int_{-\infty}^{\infty}\exp\left(-\left(\frac{|t|}{\lambda^{-1/s}}\right)^{s}\right)\d t-1\right)\\
 & =\frac{1}{s}\lambda^{-1/s}\cdot\Gamma\left(\frac{1}{s}\right)-\frac{1}{2},
\end{align}
where we have used the definition of the Gamma function $\Gamma(z)\dfn\int_{0}^{\infty}t^{z-1}e^{-t}\d t$
in the last equality.\footnote{It can also be obtained by identifying the integral as an unnormalized
generalized Gaussian distribution of zero mean, scale parameter $\lambda^{-1/s}$
and shape parameter $s$ \cite{nadarajah2005generalized}. } Further, by the convexity of $e^{-\lambda t^{s}}$ in $t\in\mathbb{\mathbb{R}}$,
\begin{align}
Z(\lambda) & \leq\frac{1}{2}\left(\int_{-\infty}^{\infty}\exp\left(-\left(\frac{|t|}{\lambda^{-1/s}}\right)^{s}\right)\d t-1+e^{-\lambda}\right).
\end{align}
Therefore
\[
Z(\lambda)=(1+a_{\lambda})\cdot\frac{1}{s}\lambda^{-1/s}\cdot\Gamma\left(\frac{1}{s}\right)
\]
where $a_{\lambda}\to0$ as $\lambda\to0$. In the same spirit, 
\begin{align}
G_{s}(P_{\lambda}) & =\sum_{i=1}^{\infty}i^{s}\cdot\frac{\exp(-\lambda i^{s})}{Z(\lambda)}\\
 & =\frac{\int_{0}^{\infty}t^{s}\exp\left(-\left(\frac{|t|}{\lambda^{-1/s}}\right)^{s}\right)\d t+b_{\lambda}}{(1+a_{\lambda})\frac{1}{s}\lambda^{-1/s}\cdot\Gamma\left(\frac{1}{s}\right)}\\
 & =\frac{\frac{1}{s}\lambda^{-\frac{s+1}{s}}\cdot\Gamma\left(\frac{s+1}{s}\right)+b_{\lambda}}{(1+a_{\lambda})\frac{1}{s}\lambda^{-1/s}\cdot\Gamma\left(\frac{1}{s}\right)}\label{eq: evaluation of guessing moment integral}\\
 & =\frac{\frac{1}{s^{2}}\lambda^{-\frac{s+1}{s}}\cdot\Gamma\left(\frac{1}{s}\right)+b_{\lambda}}{(1+a_{\lambda})\frac{1}{s}\lambda^{-1/s}\cdot\Gamma\left(\frac{1}{s}\right)}\label{eq: evaluation of guessing moment algebra}\\
 & =\frac{1}{s\lambda}(1+c_{\lambda}),\label{eq: evaluation of guessing moment asymptotics}
\end{align}
where in (\ref{eq: evaluation of guessing moment integral}), $b_{\lambda}\to0$
as $\lambda\to0$, in (\ref{eq: evaluation of guessing moment algebra})
we have used the identity $\Gamma(t+1)=t\Gamma(t)$ for $t\in\mathbb{R}_{+}$,
and in (\ref{eq: evaluation of guessing moment asymptotics}), $c_{\lambda}\to0$
as $\lambda\to0$. 

Returning to measure entropy in bits, we thus obtain that for any
distribution $P$
\[
H(P)\leq\log\left(\frac{e^{1/s}}{s}s^{1/s}\cdot G_{s}^{1/s}(P)\cdot\Gamma\left(\frac{1}{s}\right)\right)+o(1),
\]
and so 
\begin{equation}
G_{s}(P)\geq\Psi_{s}\cdot2^{sH(P)}\cdot(1+o(1)),\label{eq: entropy lower bound for P}
\end{equation}
where $\Psi_{s}\dfn e^{-1}\cdot\frac{s^{s-1}}{\Gamma^{s}(\frac{1}{s})}$
and $o(1)$ is a vanishing term as $G_{s}(P)\to\infty$. In the same
spirit, (\ref{eq: entropy lower bound for P}) holds whenever $H(P)\to\infty$. 

We return to the Boolean helper problem. Using (\ref{eq: entropy lower bound for P})
once for the guessing-moment conditioning on $f(Y^{n})=0$, and once
on $f(Y^{n})=1$ we get (see a detailed justification to (\ref{eq: entropy bound on guessing proof})
in Appendix \ref{sec: Proofs})
\begin{align}
G_{s}(X^{n}\mid f(Y^{n})) & \geq k_{n}\cdot\Psi_{s}\cdot\left[\P(f(Y^{n})=0)\cdot2^{sH(X^{n}\mid f(Y^{n})=0)}+\P(f(Y^{n})=1)\cdot2^{sH(X^{n}\mid f(Y^{n})=1)}\right]\label{eq: entropy bound on guessing proof}\\
 & \geq k_{n}\cdot\Psi_{s}\cdot2^{sH(X^{n}|f(Y^{n}))}\label{eq: entropy bound on guessing Jensen}\\
 & \geq k_{n}\cdot\Psi_{s}\cdot2^{sn-s(1-2\delta)^{2}}\label{eq: eq: entropy bound on guessing use of Boolean conjecture bound}
\end{align}
where $k_{n}\cong1$ in (\ref{eq: entropy bound on guessing proof}),
and (\ref{eq: entropy bound on guessing Jensen}) follows from Jensen's
inequality. The bound (\ref{eq: eq: entropy bound on guessing use of Boolean conjecture bound})
is directly related to the Boolean function conjecture \cite{Boolean_conjecture},
and may be proved in several ways, e.g., using Mrs. Gerber's Lemma
\cite[Th. 1]{MGL}, see \cite[Section IV]{erkip1998efficiency}\cite{Chandar_Tcham,Ordentlich_Shayevitz_Weinstein}.
\end{IEEEproof}
\begin{rem}
In \cite{massey1994guessing} the maximum-entropy problem was studied
for $s=1$. In this case, the maximum-entropy distribution is readily
identified as the geometric distribution. The proof above generalizes
that result to any $s>0$.
\end{rem}
\begin{rem}
\label{rem: If Bool conjecture true}In \cite{Ordentlich_Shayevitz_Weinstein},
the bound $H(X^{n}|f(Y^{n}))\geq n-(1-2\delta)^{2}$ used in the proof
of Theorem \ref{thm: maximum entropy} above (see (\ref{eq: eq: entropy bound on guessing use of Boolean conjecture bound}))
was improved for balanced functions, assuming $1/2(1-1/\sqrt{3})\leq\delta\leq1/2$.
Using it here leads to an immediate improvement in the bound of Theorem
\ref{thm: maximum entropy}. Furthermore, it is known \cite{Alex}
that there exists $\delta_{0}$ such that the most informative Boolean
function conjecture holds for all $\delta_{0}\leq\delta\leq1/2$.
For such crossover probabilities, 
\[
H(X^{n}|f(Y^{n}))\geq n-1+\binent(\delta)
\]
holds, and then Theorem \ref{thm: maximum entropy} may be improved
to
\begin{equation}
\gamma_{s}(\delta)\geq e^{-1}\cdot\frac{s^{s-1}\cdot(s+1)}{\Gamma^{s}(\frac{1}{s})}\cdot2^{-s\left(1-\binent(\delta)\right)}.\label{eq: imporved Max entropy is bool}
\end{equation}
\end{rem}

\subsection{A Fourier-Analytic Bound}

The second bound is based on Fourier analysis of Boolean functions
\cite{Bool_book}, and so we briefly remind the reader of the basic
definitions and results. To that end, it is convenient to assume that
the binary alphabet is $\{-1,1\}$ instead of $\{0,1\}$. An inner
product between two real-valued functions on the Boolean cube $f,g:\{-1,1\}^{n}\to\mathbb{R}$
is defined as 
\begin{equation}
\left\langle f,g\right\rangle \dfn\E\left(f(X^{n})g(X^{n})\right),\label{eq: inner product}
\end{equation}
where $X^{n}\in\{-1,1\}^{n}$ is a uniform Bernoulli vector. A \emph{character}
associated with a set of coordinates $S\subseteq[n]\dfn\{1,2,\ldots,n\}$
is the Boolean function $x^{S}\dfn\prod_{i\in S}x_{i}$, where by
convention $x^{\emptyset}=1$. It can be shown \cite[Chapter 1]{Bool_book}
that the set of all characters forms an orthonormal basis with respect
to the inner product (\ref{eq: inner product}). Furthermore, 
\[
f(x^{n})=\sum_{S\subseteq[n]}\hat{f}_{S}\cdot x^{S},
\]
where $\{\hat{f}_{S}\}_{S\subseteq[n]}$ are the \emph{Fourier coefficients
}of $f$, given by $\hat{f}_{S}=\langle x^{S},f\rangle=\E(X^{S}\cdot f(X^{n}))$.
\emph{Plancharel's identity }then states that $\langle f,g\rangle=\E(f(X^{n})g(X^{n}))=\sum_{S\in[n]}\hat{f}_{S}\hat{g}_{S}$.
The $p$ norm of a function $f$ is defined as $\Vert f\Vert_{p}\dfn[\E|f(X^{n})|^{p}]^{1/p}$. 

Letting the correlation parameter be defined as $\rho\dfn1-2\delta$,
the \emph{noise operator} is defined to be
\[
T_{\rho}f(x^{n})=\E(f(Y^{n})\mid X^{n}=x^{n}).
\]
The noise operator has a smoothing effect on the function, which is
captured by the so-called hypercontractivity theorems. Specifically,
we shall use the following version.
\begin{thm}[{\cite[p. 248]{Bool_book}}]
\label{thm: (p,2) hyper} Let $f:\{-1,1\}^{n}\to\mathbb{R}$ and
$0\leq\rho\leq1$. Then, $\Vert T_{\rho}f\Vert_{2}\leq\Vert f\Vert_{\rho^{2}+1}$.
\end{thm}
Our Fourier-based bound is as follows:
\begin{thm}
\label{thm: Fourier}We have
\[
\gamma_{s}(\delta)\geq\max_{0\leq\lambda\leq1}\left[1-\frac{(s+1)\cdot(1-2\delta)^{\lambda}}{\left[(1+(1-2\delta)^{2(1-\lambda)})s+1\right]^{\frac{1}{(1+(1-2\delta)^{2(1-\lambda)})}}}\right].
\]
\end{thm}
This bound can be weakened by the possibly sub-optimal choice $\lambda=1$,
which leads to the simpler and explicit bound:
\begin{cor}
We have
\begin{equation}
\gamma_{s}(\delta)\geq1-\frac{(s+1)\cdot(1-2\delta)}{\sqrt{1+2s}}.\label{eq: Simple Fourier bound}
\end{equation}
\end{cor}
\begin{IEEEproof}[Proof of Theorem \ref{thm: Fourier}]
From Bayes law (recall that $X_{i}\in\{-1,1\}$)
\[
\P(X^{n}=x^{n}\mid f(Y^{n})=b)=2^{-(n+1)}\cdot\frac{1+bT_{\rho}f(x^{n})}{\P(f(Y^{n})=b)},
\]
and from the law of total expectation
\begin{equation}
G_{s}(X^{n}\mid f(Y^{n}))=\P(f(Y^{n})=1)\cdot G_{s}(X^{n}\mid f(Y^{n})=1)+\P(f(Y^{n})=-1)\cdot G_{s}(X^{n}\mid f(Y^{n})=-1).\label{eq: gussing moment - total expectation pm 1}
\end{equation}

Letting $\hat{f}_{\phi}=\E f(X^{n})$, and defining $g=f-\hat{f}_{\phi}$,
the first addend on the r.h.s. of (\ref{eq: gussing moment - total expectation pm 1})
is given by
\begin{align}
\P(f(Y^{n})=1)\cdot G_{s}(X^{n}\mid f(Y^{n})=1) & =2^{-(n+1)}\sum_{x^{n}}\left(1+\hat{f}_{\phi}+T_{\rho}g(x^{n})\right)\cdot\ord_{T_{\rho}g}^{s}(x^{n})\\
 & =\frac{(1+\hat{f}_{\phi})}{2}\cdot\E\left(\ord_{T_{\rho}g}^{s}(X^{n})\right)+\frac{1}{2}\langle T_{\rho}g,\ord_{T_{\rho}g}^{s}\rangle\\
 & =\frac{(1+\hat{f}_{\phi})}{2}\cdot K_{s}(2^{n})+\frac{1}{2}\langle T_{\rho}g,\ord_{T_{\rho}g}^{s}\rangle\\
 & =\frac{(1+\hat{f}_{\phi})}{2}\cdot\ell_{n}\cdot\frac{2^{sn}}{s+1}+\frac{1}{2}\langle T_{\rho}g,\ord_{T_{\rho}g}^{s}\rangle,\label{eq: guessing - condition on f=00003D1 with inner product}
\end{align}
where in the last equality, $\ell_{n}\cong1$ (Lemma \ref{lem: moments of ordinal}).
Let $\lambda\in[0,1]$, and denote $\rho_{1}\dfn\rho^{\lambda}$ and
$\rho_{2}=\rho^{1-\lambda}$. Then, the inner-product term in (\ref{eq: guessing - condition on f=00003D1 with inner product})
may be upper bounded as
\begin{align}
\left|\langle T_{\rho}g,\ord_{T_{\rho}g}^{s}\rangle\right| & =\left|\langle T_{\rho_{1}}g,T_{\rho_{2}}\ord_{T_{\rho}g}^{s}\rangle\right|\label{eq: bounding inner product adjoint}\\
 & \leq\Vert T_{\rho_{1}}g\Vert_{2}\cdot\Vert T_{\rho_{2}}\ord_{T_{\rho}g}^{s}\Vert_{2}\label{eq: bounding inner product CS}\\
 & =\rho_{1}\cdot\sqrt{1-\hat{f}_{\phi}^{2}}\cdot\Vert T_{\rho_{2}}\ord_{T_{\rho}g}^{s}\Vert_{2}\label{eq: bounding inner product derivation}\\
 & \leq\rho_{1}\cdot\sqrt{1-\hat{f}_{\phi}^{2}}\cdot\Vert\ord_{T_{\rho}g}^{s}\Vert_{1+\rho_{2}^{2}}\label{eq: bounding inner product hyper}\\
 & =\rho_{1}\cdot\sqrt{1-\hat{f}_{\phi}^{2}}\cdot\left(K_{(1+\rho_{2}^{2})s}(2^{n})\right)^{1/(1+\rho_{2}^{2})}\\
 & =\rho_{1}\cdot\sqrt{1-\hat{f}_{\phi}^{2}}\cdot\left[k_{n}\cdot\frac{1}{(1+\rho_{2}^{2})s+1}\right]^{1/(1+\rho_{2}^{2})}\cdot2^{sn},\label{eq: bounding inner product}
\end{align}
where (\ref{eq: bounding inner product adjoint}) holds since $T_{\rho}$
is a self-adjoint operator, (\ref{eq: bounding inner product CS})
follows from the Cauchy\textendash Schwarz inequality. For (\ref{eq: bounding inner product derivation})
note that 
\begin{align}
\Vert T_{\rho}g\Vert_{2}^{2} & =\langle T_{\rho}g,T_{\rho}g\rangle\\
 & =\sum_{S\in[n]}\rho^{2|S|}\hat{g}_{S}^{2}\label{eq: noise operator g norm Fourier}\\
 & =\sum_{S\in[n]\backslash\phi}\rho^{2|S|}\hat{f}_{S}^{2}\label{eq: noise operator f norm Fourier}\\
 & \leq\rho^{2}\cdot(1-\hat{f}_{\phi}^{2}),\label{eq: noise operator g norm bound}
\end{align}
where (\ref{eq: noise operator g norm Fourier}) follows from Plancharel's
identity, (\ref{eq: noise operator f norm Fourier}) is since $\hat{g}_{S}=\hat{f}_{S}$
for all $S\neq\phi$, and $\hat{g}_{\phi}=0$, and (\ref{eq: noise operator g norm bound})
follows from $\sum_{S\in[n]}\hat{f}_{S}^{2}=\Vert f\Vert_{2}^{2}=\E f^{2}=1$.
Equation (\ref{eq: bounding inner product hyper}) follows from Theorem
\ref{thm: (p,2) hyper}, and in (\ref{eq: bounding inner product}),
$k_{n}\cong1$. The second addend on the r.h.s. of (\ref{eq: gussing moment - total expectation pm 1})
can be bounded in the same manner. Hence, 
\begin{align}
G_{s}(X^{n}\mid f(Y^{n})) & \geq\max_{0\leq\lambda\leq1}2^{sn}\cdot\left[\ell_{n}\cdot\frac{1}{s+1}-\rho^{\lambda}\cdot\sqrt{1-\hat{f}_{\phi}^{2}}\cdot\left[k_{n}\cdot\frac{1}{(1+\rho^{2(1-\lambda)})s+1}\right]^{1/(1+\rho^{2(1-\lambda)})}\right]\\
 & \geq\max_{0\leq\lambda\leq1}2^{sn}\cdot\left[\ell_{n}\cdot\frac{1}{s+1}-\rho^{\lambda}\left[k_{n}\cdot\frac{1}{(1+\rho^{2(1-\lambda)})s+1}\right]^{1/(1+\rho^{2(1-\lambda)})}\right]\\
 & \to2^{sn}\cdot\max_{0\leq\lambda\leq1}\left[\frac{1}{s+1}-\frac{\rho^{\lambda}}{\left[(1+\rho^{2(1-\lambda)})s+1\right]^{1/(1+\rho^{2(1-\lambda)})}}\right]
\end{align}
as $n\to\infty$. 
\end{IEEEproof}

\section{Graphs\label{sec: Graphs}}

\begin{figure}
\begin{centering}
\includegraphics[scale=0.5]{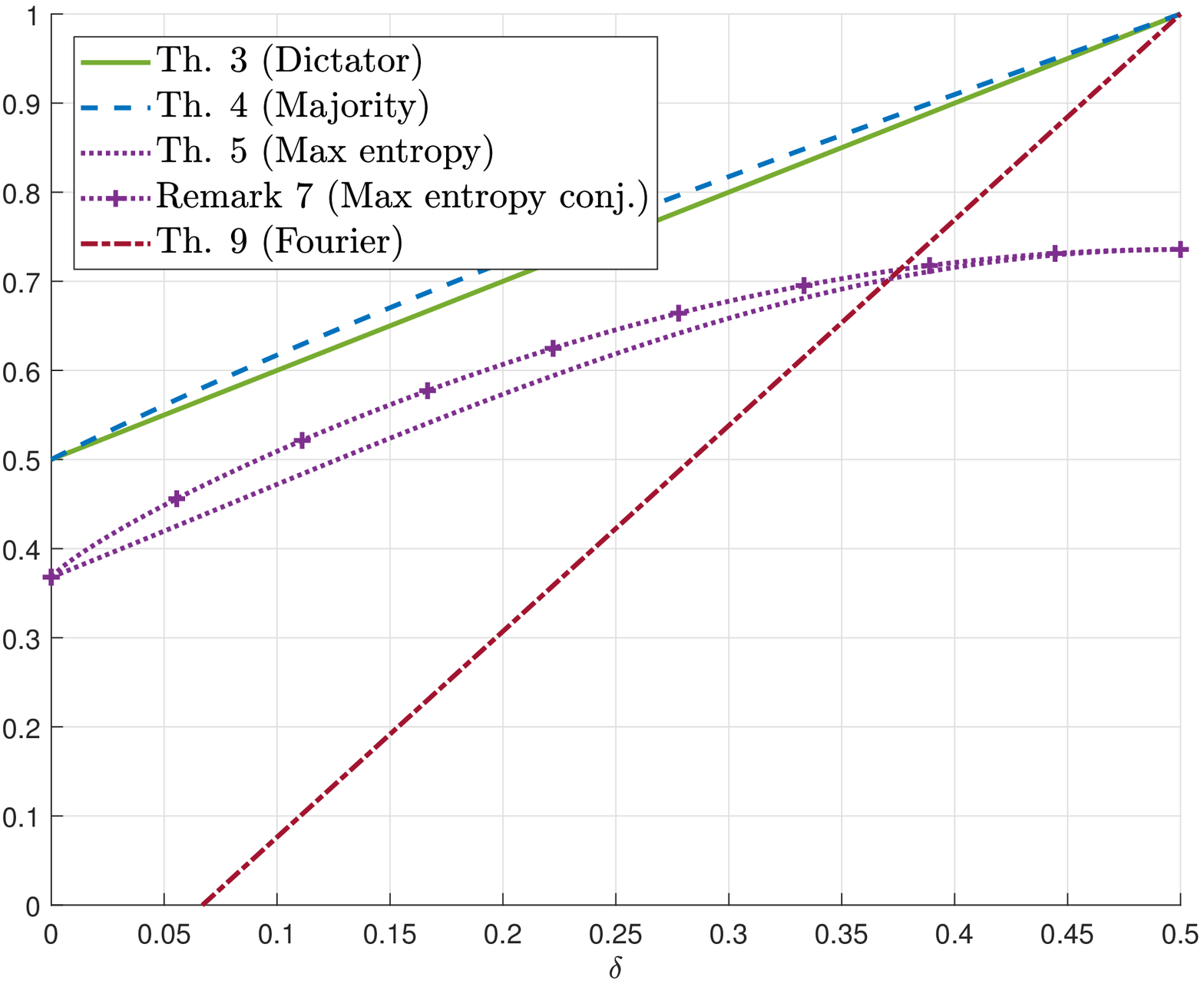}\includegraphics[scale=0.5]{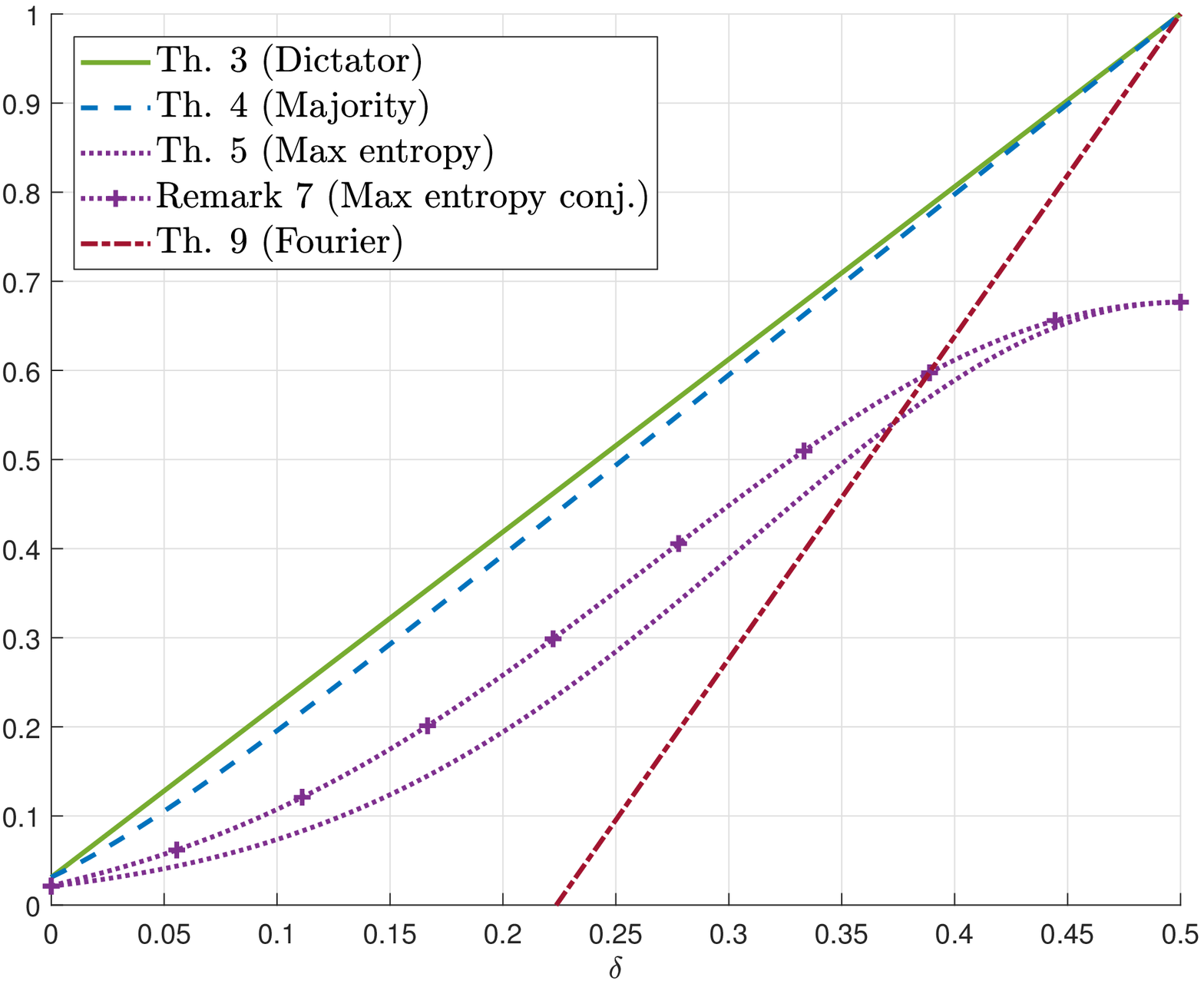}
\par\end{centering}
\caption{Bounds on $\gamma_{s}(\delta)$ for $s=1$ (left) and $s=5$ (right)
and varying $\delta\in[0,1/2]$.\label{fig: fixed rho}}

\end{figure}

\begin{figure}
\begin{centering}
\includegraphics[scale=0.5]{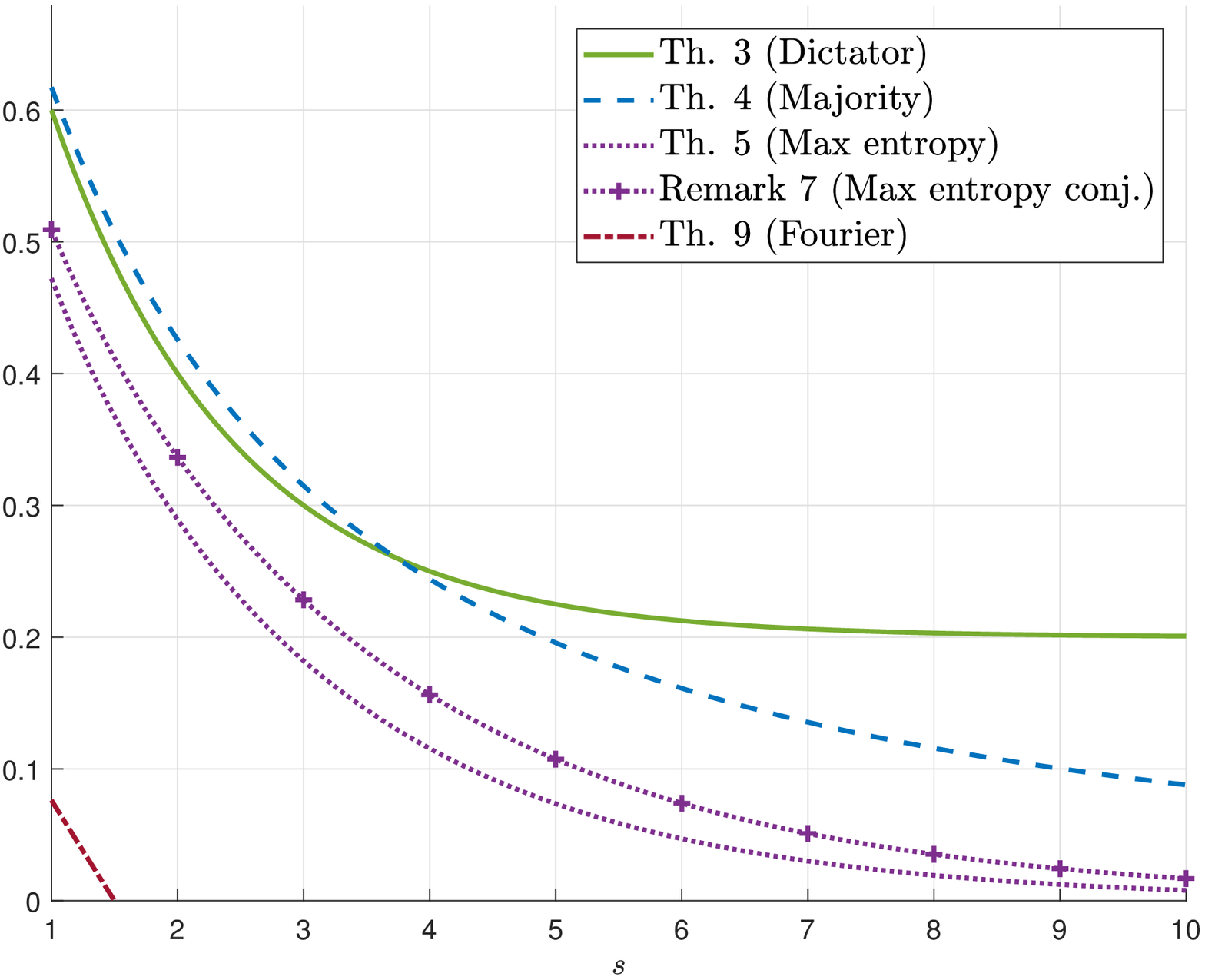}\includegraphics[scale=0.5]{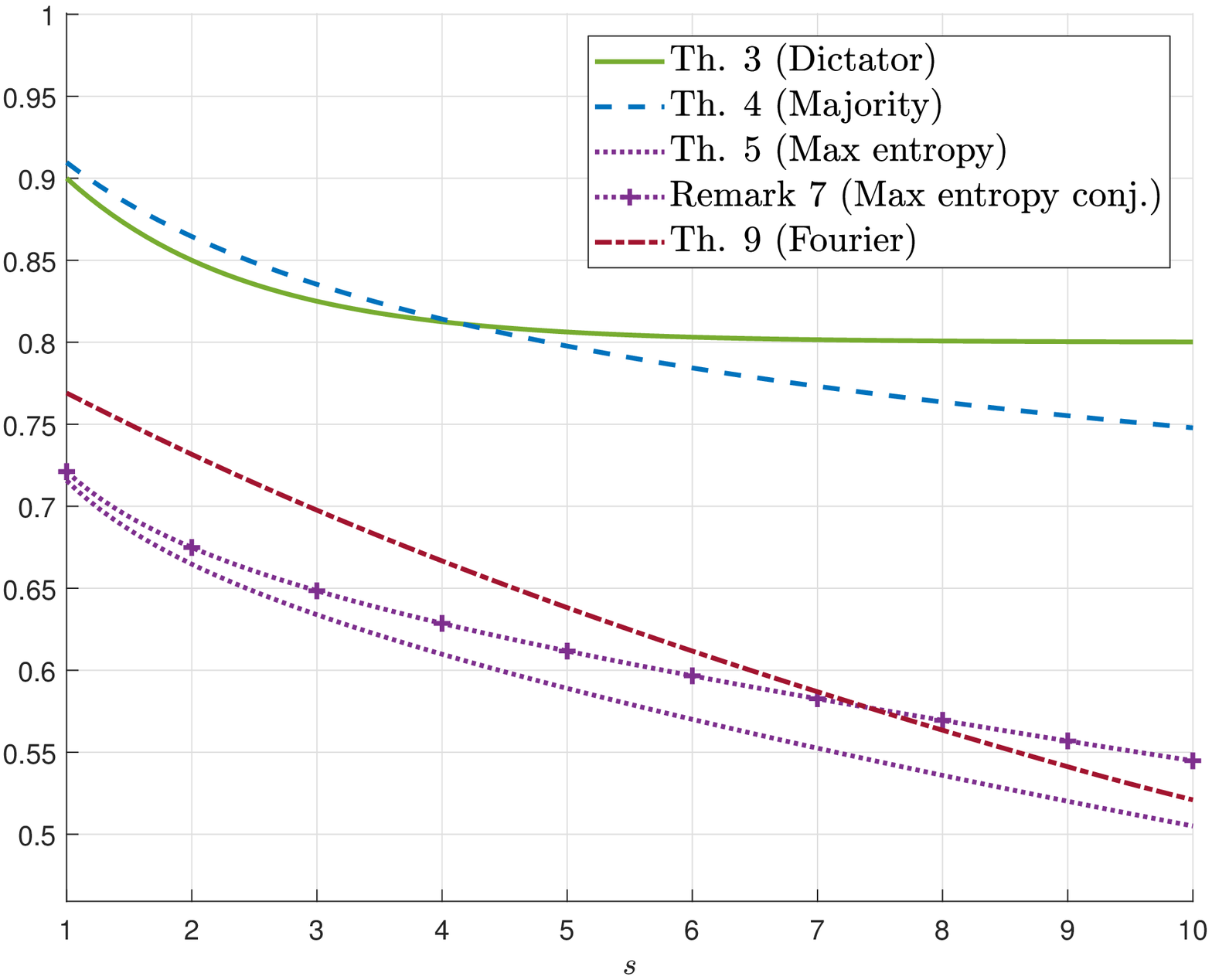}
\par\end{centering}
\caption{Bounds on $\gamma_{s}(\delta)$ for $\delta=0.1$ (left) and $\delta=0.4$
(right) and varying $s\in[1,10]$.\label{fig: fixed p}}
\end{figure}

In Fig. \ref{fig: fixed rho} (resp. Fig. \ref{fig: fixed p}) the
bounds on $\gamma_{s}(\delta)$ are plotted for fixed values of $s$
(resp. $\delta$). As for upper bounds, it can be found that when
$s\lesssim3.5$ Dictator dominates Majority (for all values of $\delta$),
whereas for $s\gtrsim4.25$ Majority dominates Dictator. For $3.5\lesssim s\lesssim4.25$
there exists $\delta'_{s}$ such that Majority is better for $\delta\in(0,\delta'_{s})$
and Dictator is better for $\delta\in(\delta'_{s},1/2)$. Fig. \ref{fig: fixed p}
demonstrates the switch from Dictator to Majority as $s$ increases
(depending on $\delta$). As for lower bounds, first note that the
conjectured maximum-entropy bound (\ref{eq: imporved Max entropy is bool})
was also plotted (see Remark \ref{rem: If Bool conjecture true}).
It can be observed that the maximum-entropy bound is better for low
values of $\delta$, whereas the Fourier analysis bound is better
for high values. As a function of $s$, it seems that the maximum-entropy
bound (resp. Fourier-analysis bound) is better for high (resp. low)
values of $s$. Finally, we mention that the maximizing parameter
in the Fourier-based bound (Theorem \ref{thm: Fourier}) is $\lambda=1$,
and the resulting bound is as in (\ref{eq: Simple Fourier bound}).
For values of $s$ as low as $10$, the maximizing $\lambda$ may
be far from $1$, and in fact it continuously and monotonically increases
from $0$ to $1$ as $\delta$ increases from $0$ to $1/2$. 

\section{Guessing Efficiency for a General Binary Input Channel \label{sec: General source-channel}}

In this section, we consider the guessing efficiency for general channels
with a uniform binary input. The lower bound of Theorem \ref{thm: maximum entropy}
can be easily generalized for this case. To that end, consider the
SDPI constant \cite{ahlswede1976spreading,raginsky2016strong} of
the reverse channel $(P_{Y},P_{X|Y})$, given by 
\begin{equation}
\eta(P_{Y},P_{X|Y})\dfn\sup_{Q_{Y}\colon Q_{Y}\neq P_{Y}}\frac{D(Q_{X}||P_{X})}{D(Q_{Y}||P_{Y})},\label{eq: SDPI constant}
\end{equation}
where $Q_{X}$ is the $X$-marginal of $Q_{Y}\circ P_{X|Y}$ . As
was shown in \cite[Th. 2]{anantharam2014hypercontractivity}, the
SDPI constant of $(P_{Y},P_{X|Y})$ is also given by 
\begin{equation}
\eta(P_{Y},P_{Y|X})=\sup_{P_{W|Y}\colon W-Y-X,\;I(W;Y)>0}\frac{I(W;X)}{I(W;Y)}.\label{eq: SDPI as MI ratio}
\end{equation}

\begin{thm}
\label{thm: maximum entropy general}We have\textbf{ } 
\begin{equation}
\gamma_{s}(P_{X},P_{Y|X})\geq e^{-1}\cdot\frac{s^{s-1}\cdot(s+1)}{\Gamma^{s}(\frac{1}{s})}\cdot2^{-s\cdot\eta(P_{Y},P_{Y|X})}.\label{eq: entropy lower bound general}
\end{equation}
\end{thm}
\begin{IEEEproof}
The proof follows the same lines of the proof of Theorem \ref{thm: maximum entropy},
up to (\ref{eq: entropy bound on guessing Jensen}), yielding 
\begin{equation}
G_{s}(X^{n}\mid f(Y^{n}))\geq k_{n}\cdot\Psi_{s}\cdot2^{s\left[n-I(X^{n};f(Y^{n}))\right]}.\label{eq: entropy lower bound with mutual information}
\end{equation}
Now, let $W^{(n)}$ be such that $X^{n}-Y^{n}-W^{(n)}$ forms a Markov
chain. Then, 
\begin{align}
\sup_{f\colon{\cal Y}^{n}\to\{0,1\}}\frac{I(X^{n};f(Y^{n}))}{I(Y^{n};f(Y^{n}))} & \leq\sup_{P_{W^{(n)}|Y^{n}}}\frac{I(X^{n};W^{(n)})}{I(Y^{n};W^{(n)})}\\
 & =\eta(P_{Y^{n}},P_{X^{n}|Y^{n}})\\
 & =\eta(P_{Y},P_{X|Y}),\label{eq: SDPI tensorizes}
\end{align}
where (\ref{eq: SDPI tensorizes}) follows since the SDPI constant
tensorizes (see \cite{anantharam2014hypercontractivity} for an argument
obtained by relating the SDPI constant to the hypercontractivity parameter,
or \cite[p. 5]{anantharam2013maximal} for a direct proof). Thus,
for all $f$ 
\begin{align}
I(X^{n};f(Y^{n})) & \leq\eta(P_{Y},P_{X|Y})\cdot I(Y^{n};f(Y^{n}))\\
 & \leq\eta(P_{Y},P_{X|Y})\cdot H(f(Y^{n}))\label{eq: SDPI and entropy}\\
 & \leq\eta(P_{Y},P_{X|Y}).\label{eq: SDPI bound}
\end{align}
Inserting (\ref{eq: SDPI bound}) to (\ref{eq: entropy lower bound with mutual information}).
Therefore 
\[
G_{s}(X^{n}\mid f(Y^{n}))\geq k_{n}\cdot\Psi_{s}\cdot2^{s\left[n-\eta(P_{Y},P_{X|Y})\right]},
\]
and using the definition of the guessing efficiency (\ref{eq: guessing efficiency optimal})
completes the proof. 
\end{IEEEproof}
\begin{rem}
It is evident from (\ref{eq: SDPI and entropy}) that if the helper
is allowed to send $k$ bits, then the associated $k$-bit guessing
efficiency is lower bounded by 
\[
\gamma_{s}^{(k)}(\delta)\geq e^{-1}\cdot\frac{s^{s-1}\cdot(s+1)}{\Gamma^{s}(\frac{1}{s})}\cdot2^{-s\cdot k\cdot\eta(P_{Y},P_{Y|X})}.
\]
\end{rem}
\begin{rem}
The bound for the BSC case (Theorem \ref{thm: maximum entropy}) is
indeed a special case of Theorem \ref{thm: maximum entropy general}
as the reverse BSC channel is also a BSC with uniform input and the
same crossover probability. For BSCs, it is well known that the SDPI
constant is $(1-2\delta)^{2}$ \cite[Th. 9]{ahlswede1976spreading}.

Next, we consider in more detail the case where the observation channel
is a BEC. 
\end{rem}

\subsection{Binary Erasure Channel}

Suppose that $Y^{n}\in\{0,1,\e\}^{n}$ is obtained from $X^{n}$ by
erasing each bit independently with probability $\epsilon\in[0,1]$.
As before, Bob observes the channel output $Y^{n}$ and can send one
bit $f:\{0,1,\e\}^{n}\to\{0,1\}$ to Alice, who wishes to guess $X^{n}$.
With a slight abuse of notation, the guessing efficiency (\ref{eq: guessing efficiency optimal})
will be denoted by $\gamma_{s}(\epsilon)$.

\begin{figure}
\begin{centering}
\includegraphics[scale=0.5]{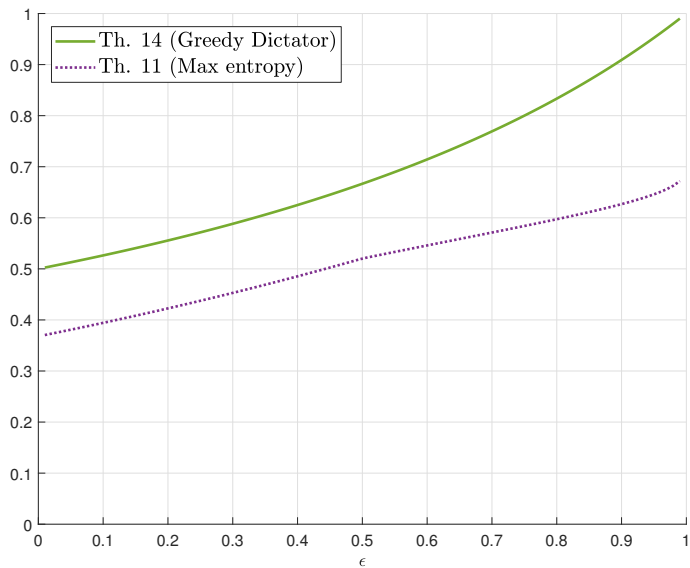}
\par\end{centering}
\caption{Bounds on $\gamma_{s}(\delta)$ for $s=1$ varying $\epsilon\in[0,1]$.\label{fig: BEC fixed rho}}
\end{figure}

To compute the lower bound of Theorem \ref{thm: maximum entropy general},
we need to find the SDPI constant associated with the reverse channel,
which is easily verified to be
\[
P_{X|Y=y}=\begin{cases}
y, & y=0\text{ or }y=1\\
\text{Ber}(1/2), & y=\e
\end{cases},
\]
with an input distribution $P_{Y}=(\frac{1-\epsilon}{2},\epsilon,\frac{1-\epsilon}{2})$.
Letting $Q_{Y}(y)=q_{y}$ for $y\in\{0,1,\e\}$ yields $Q_{X}(x)=q_{x}+\frac{q_{\e}}{2}$
for $x\in\{0,1\}$. The computation of $\eta(P_{Y},P_{X|Y})$ is now
a simple three-dimensional constrained optimization problem. We plotted
the resulting lower bound for $s=1$ in Fig. \ref{fig: BEC fixed rho}.

Let us now turn to upper bounds, and focus for simplicity on the average
guessing time, i.e., the guessing-moment for $s=1$. To begin, let
$S$ represent the set of indices of the symbols that were not erased,
i.e., $i\in S$ if and only if $Y_{i}\neq\e$. Any function $f:\{0,1,\e\}^{n}\to\{0,1\}$
is then uniquely associated with a set of Boolean functions $\{f_{S}\}_{S\in[n]}$,
where $f_{S}:\{0,1\}^{|S|}\to\{0,1\}$ designates the operation of
the function when $S$ is the set of non-erased symbols. We also let
$\Pr(S)=(1-\epsilon)^{|S|}\cdot\epsilon^{|S^{c}|}$ be the probability
that the non-erased symbols have index set $S$. Then, the joint probability
distribution is given by 
\begin{align}
\P(X^{n}=x^{n},f(Y^{n})=1) & =\P(X^{n}=x^{n})\cdot\P(f(Y^{n})=1\mid X^{n}=x^{n})\\
 & =2^{-n}\cdot\sum_{S\subseteq[n]}\Pr(S)\cdot\P(f(Y^{n})=1\mid X^{n}=x^{n},S)\\
 & =2^{-n}\cdot\sum_{S\subseteq[n]}\Pr(S)\cdot f_{S}(x^{n}),\label{eq: posterior over BEC 1}
\end{align}
and, similarly, 
\begin{align}
\P(X^{n}=x^{n},f(Y^{n})=0) & =2^{-n}\cdot\sum_{S\subseteq[n]}\Pr(S)\cdot(1-f_{S}(x^{n}))\\
 & =2^{-n}-2^{-n}\cdot\sum_{S\subseteq[n]}\Pr(S)\cdot f_{S}(x^{n}).\label{eq: posterior over BEC 0}
\end{align}
Interestingly, for any given $f$, the optimal guessing order given
that $f(Y^{n})=0$ is reversed to the optimal guessing order when
$f(Y^{n})=1$. Also apparent is that the posterior probability is
determined by a mixture of $2^{n}$ different Boolean functions $\{f_{S}\}_{S\in[n]}$.
This may be contrasted with the BSC case, in which the posterior is
determined by a \emph{single} Boolean function (though with noisy
input). 

A seemingly natural choice is a \emph{Greedy Dictator }function\emph{,}
for which $f(Y^{n})$ sends the first non-erased bit. Concretely,
letting 
\[
k(y^{n})\dfn\begin{cases}
n+1, & y^{n}=\e^{n}\\
\min\left\{ i\colon y_{i}\neq\e\right\} , & \text{otherwise}
\end{cases},
\]
the \emph{Greedy Dictator} is defined by 
\begin{equation}
\gdic(y^{n})\dfn\begin{cases}
\text{Ber}(1/2), & y^{n}=\e^{n}\\
y_{k(y^{n})}, & \text{otherwise}
\end{cases},\label{eq: greedy dictator}
\end{equation}
where $\text{Ber}(\alpha)$ is a Bernoulli r.v. of success probability
$\alpha$. From an analysis of the posterior probability (see Appendix
\ref{sec: Proofs}), it is evident that conditioned on $f(Y^{n})=0$,
an optimal guessing order must satisfy that $x^{n}$ is guessed before
$z^{n}$ whenever 
\begin{equation}
\sum_{i=1}^{n}\epsilon^{i-1}\cdot x_{i}\leq\sum_{i=1}^{n}\epsilon^{i-1}\cdot z_{i},\label{eq: greedy dictator gussing rule}
\end{equation}
which can be loosely thought of as comparing the \textquotedbl{}base
$1/\epsilon$ expansion\textquotedbl{} of $x^{n}$ and $z^{n}$. Furthermore,
when $\epsilon$ is close to $1$, then the optimal guessing order
tends toward a \emph{minimum Hamming weight} rule (of maximum in case
$f=1$).

The Greedy Dictator function is ``local optimal'' when $\epsilon\in[0,1/2]$,
in the following sense: 
\begin{prop}
\label{prop: local minima of gdic}If $\epsilon\in[0,1/2]$ then an
optimal guessing order conditioning on $\gdic(Y^{n})=0$ (resp. $\gdic(Y^{n})=1$)
is lexicographic (reverse lexicographic). Also, given lexicographic
(reverse lexicographic) order when the received bit is $0$ (resp.
$1$), the optimal function $f$ is Greedy Dictator. 
\end{prop}
\begin{IEEEproof}
See Appendix \ref{sec: Proofs}. 
\end{IEEEproof}
The guessing efficiency of the Greedy Dictator for $s=1$ can be evaluated,
and the analysis leads to the following upper bound: 
\begin{thm}
\label{thm: Greedy dictator}We have 
\[
\gamma_{1}(\epsilon)\leq\frac{1}{2-\epsilon},
\]
and the r.h.s. above is achieved with equality by the Greedy Dictator
for $\epsilon\in[0,1/2]$. 
\end{thm}
\begin{IEEEproof}
See Appendix \ref{sec: Proofs}. 
\end{IEEEproof}
The upper bound of Theorem \ref{thm: Greedy dictator} is plotted
in Fig. \ref{fig: BEC fixed rho}. Based on Proposition \ref{prop: local minima of gdic}
and numerical computations for moderate values of $n$ we conjecture: 
\begin{conjecture}
\label{conj:Greedy-dictator}Greedy Dictator functions attain $\gamma_{s}(\epsilon)$
for the BEC. 
\end{conjecture}
Supporting evidence for this conjecture includes the local optimality
of Proposition \ref{prop: local minima of gdic} (although there are
other locally optimal choices), as well as the following heuristic
argument: Intuitively, Bob should reveal as much as possible regarding
the bits he has seen and as little as possible regarding the erasure
pattern. So, it seems reasonable to find a smallest possible set of
balanced functions from which to choose all the functions $f_{S}$,
so that they coincide as much as possible. Greedy Dictator is a greedy
solution to this problem: it uses the function $x_{1}$ for half of
the erasure patterns, which is the maximum possible. Then, it uses
the function $x_{2}$ for half of the remaining patterns, and so on.
Indeed, we were not able to find a better function than $\gdic$ for
small values of $n$.

However, applying standard techniques in attempt to prove Conjecture
\ref{conj:Greedy-dictator} has not been fruitful. One possible technique
is induction. For example, assume that the optimal functions for dimension
$n-1$ are $f_{S}^{(n-1)}$. Then, it might be perceived that there
exists a bit, say $x_{1}$, such that the optimal functions for dimension
$n$ satisfy $f_{S}^{(n)}=f_{S}^{(n-1)}$ if $x_{1}$ is erased; in
that case, it remains only to determine $f_{S}^{(n)}$ when $x_{1}$
is not erased. However, observing (\ref{eq: posterior over BEC 1}),
it is apparent that the optimal choice of $f_{S}^{(n)}$ should satisfy
two contradicting goals \textendash{} on the one hand, to match the
order induced by 
\begin{equation}
\sum_{S\subseteq[n]\colon1\notin S}\Pr(S)\cdot f_{S}(x^{n})\label{eq: posterior - one is erased}
\end{equation}
and on the other hand, to minimize the average guessing time of 
\begin{equation}
\sum_{S\subseteq[n]\colon1\in S}\Pr(S)\cdot f_{S}(x^{n}).\label{eq: posterior - one is not erased}
\end{equation}
It is easy to see that taking a greedy approach toward satisfying
the second goal would result in $f_{S}^{(n)}(x^{n})=x_{1}$ if $1\in S$,
and performing the recursion steps would indeed lead to a Greedy Dictator
function. Interestingly, taking a greedy approach toward satisfying
the first goal would also lead to a Greedy Dictator function, but
one which operates on a cyclic permutation of the inputs (specifically,
(\ref{eq: greedy dictator}) applied to $(y_{2}^{n},y_{1})$). Nonetheless,
it is not clear that choosing $\{f_{S}^{(n)}\}_{S\colon1\in S}$ with
some loss in the average guessing time induced by (\ref{eq: posterior - one is not erased})
could not lead to a gain in the second goal (matching the order of
(\ref{eq: posterior - one is erased})) which outweighs that loss.

Another possible technique is majorization. It is well known that
if one probability distribution majorizes another, then all the non-negative
guessing-moments of the first are no greater than the corresponding
moments of the second \cite[Proposition 1]{burin2017reducing}.\footnote{The proof in \cite{burin2017reducing} is only for $s=1$, but it
is easily extended to the general $s>0$ case. } Hence, one approach toward identifying the optimal function could
be to try and find a function whose induced posterior distributions
majorize the corresponding posteriors induces by any other functions
with the same bias (it is of course not clear that such a function
even exists). This approach unfortunately fails for the Greedy Dictator.
For example, the posterior distributions induced by setting $f_{S}$
to be Majority functions are not always majorized by those induces
by the Greedy Dictator (although they seem to be \textquotedbl{}almost\textquotedbl{}
majorized), e.g. for $n=5$ and $\epsilon=0.4$, even though the average
guessing time of Greedy Dictator is lower. In fact, the guessing moments
for Greedy Dictator seem to be better than these of Majority irrespective
of the value of $s$.

\section*{Acknowledgments}

We are very grateful to Amos Lapidoth and Robert Graczyk for discussing
their recent work on guessing with a helper \cite{Graczyk_Lapidoth,gr:17:thesis}
during the second author's visit to ETH, which provided the impetus
for this work. 

\appendices{}

\section{Proofs \label{sec: Proofs}}
\begin{IEEEproof}[Proof of Proposition \ref{prop: general properties} ]
 The claim that random functions do not improve beyond deterministic
ones follows directly from that property that conditioning reduces
guessing-moment \cite[Corollary 1]{arikan1996inequality}. Monotonicity
follows from the fact that Bob can always simulate a noisier channel.
Now, if $\delta=1/2$ then $X^{n}$ and $Y^{n}$ are independent,
and $G_{s}(X^{n}\mid f(Y^{n}))=G_{s}(X^{n})\cong\frac{2^{sn}}{s+1}$
for any $f$ (Lemma \ref{lem: moments of ordinal}). For $\delta=0$,
let $\P(f(Y^{n})=1)\dfn q$, and assume without loss of generality
that $q\leq1/2$. Then,
\begin{align}
G_{s}(X^{n}\mid f(Y^{n})) & =(1-q)\cdot K_{s}((1-q)\cdot2^{n})+q\cdot K_{s}(q\cdot2^{n})\\
 & =2^{-n}\cdot\sum_{i=1}^{(1-q)\cdot2^{n}}i^{s}+2^{-n}\cdot\sum_{i=1}^{q\cdot2^{n}}i^{s}\\
 & =2^{-n}\cdot\left[\sum_{i=1}^{2^{n-1}}i^{s}+\sum_{i=2^{n-1}+1}^{(1-q)\cdot2^{n}}i^{s}+\sum_{i=1}^{2^{n-1}}i^{s}-\sum_{i=q\cdot2^{n}}^{2^{n-1}}i^{s}\right]\\
 & \geq2^{-(n-1)}\sum_{i=1}^{2^{n-1}}i^{s}\\
 & =K_{s}(2^{n-1}),
\end{align}
with equality when $q=1/2$. Thus, the minimal $G_{s}(X^{n}\mid f(Y^{n}))$
is obtained by any balanced function, and equal to $K_{s}(2^{n-1})\cong\frac{2^{s(n-1)}}{s+1}$
(Lemma \ref{lem: moments of ordinal}).

To prove that the limit in (\ref{eq: guessing efficiency optimal})
exists, first note that 
\begin{align}
G_{s}(X^{n+1}) & =2^{-(n+1)}\sum_{i=1}^{2^{n+1}}i^{s}\\
 & =2^{-(n+1)}\sum_{i=1}^{2^{n}}(2i-1)^{s}+(2i)^{s}\\
 & \geq2^{s}\cdot2^{-n}\sum_{i=1}^{2^{n}}(i-1)^{s}\\
 & =\ell_{n}\cdot2^{s}\cdot2^{-n}\sum_{i=1}^{2^{n}}i^{s}\\
 & =\ell_{n}\cdot2^{s}\cdot G_{s}(X^{n}),
\end{align}
where
\begin{equation}
\ell_{n}\dfn\frac{\sum_{i=1}^{2^{n}}(i-1)^{s}}{\sum_{i=1}^{2^{n}}i^{s}}.\label{eq: defintion of ration close sums}
\end{equation}
Second, let 
\[
\gamma_{s}^{(n)}(\delta)\dfn\min_{f:\{0,1\}^{n}\to\{0,1\}}\frac{G_{s}(X^{n}\mid f(Y^{n}))}{G_{s}(X^{n})},
\]
and let $\{f_{n}^{*}\}$ be a sequence of functions such that $f_{n}^{*}$
achieves $\gamma_{s}^{(n)}(\delta)$. Denote the order induced by
the posterior $\P(X^{n}=x^{n}\mid f_{n}^{*}(Y^{n})=b)$ as $\ord_{b,n,n}$,
$b\in\{0,1\}$, and the order induced by $\P(X^{n+1}=x^{n+1}\mid f_{n}^{*}(Y^{n})=b)$
as $\ord_{b,n,n+1}$. As before (when breaking ties arbitrarily)
\[
\ord_{b,n,n+1}(x^{n},0)=2\ord_{b,n,n}(x^{n})
\]
and
\[
\ord_{b,n,n+1}(x^{n},1)=2\ord_{b,n,n}(x^{n})-1\leq2\ord_{b,n,n}(x^{n}).
\]
Thus,
\begin{align}
 & G_{s}(X^{n+1}\mid f_{n+1}^{*}(Y^{n+1}))\nonumber \\
 & \leq G_{s}(X^{n+1}\mid f_{n}^{*}(Y^{n}))\\
 & =\P(f_{n}^{*}(Y^{n+1})=0)\cdot G_{s}(X^{n+1}\mid f_{n}^{*}(Y^{n})=0)+\P(f_{n}^{*}(Y^{n})=1)\cdot G_{s}(X^{n+1}\mid f_{n}^{*}(Y^{n})=1)\\
 & \leq\sum_{x^{n+1}}\P(X^{n+1}=x^{n+1},f_{n}^{*}(Y^{n})=0)\cdot\ord_{0,n,n+1}^{s}(x^{n+1})\nonumber \\
 & \hphantom{=}+\sum_{x^{n+1}}\P(X^{n+1}=x^{n+1},f_{n}^{*}(Y^{n})=1)\cdot\ord_{1,n,n+1}^{s}(x^{n+1})\\
 & \leq2^{s}\cdot\sum_{x^{n}}\P(X^{n}=x^{n},f_{n}^{*}(Y^{n})=0)\cdot\ord_{0,n,n}^{s}(x^{n})+\P(X^{n}=x^{n},f_{n}^{*}(Y^{n})=1)\cdot\ord_{1,n,n}^{s}(x^{n})\\
 & =2^{s}\cdot G_{s}(X^{n}\mid f_{n}^{*}(Y^{n})).
\end{align}
Hence,
\begin{equation}
\gamma_{s}^{(n+1)}(\delta)\leq\ell_{n}^{-1}\cdot\gamma_{s}^{(n)}(\delta).\label{eq: approximate monotoncity}
\end{equation}
To continue, we further explore $\ell_{n}$. Noting that we can start
the summation in the numerator (\ref{eq: defintion of ration close sums})
from $i=2$, and using (\ref{eq: sums using integral upper bound})
and (\ref{eq: sums using integral lower bound}) (proof of Lemma \ref{lem: moments of ordinal}
below), we get 
\begin{align}
1 & \geq\ell_{n}\\
 & \geq\frac{\frac{1}{s+1}\cdot\frac{2^{n(s+1)}-1}{2^{n}-1}}{\frac{1}{s+1}\cdot\frac{(2^{n}+1)^{s+1}-1}{2^{n}}}\\
 & \geq\frac{2^{n(s+1)}-1}{(2^{n}+1)^{s+1}}\\
 & =\left(\frac{2^{n}}{2^{n}+1}\right)^{s+1}-\frac{1}{2^{n(s+1)}}\\
 & =\left(1+\frac{1}{2^{n}}\right)^{-(s+1)}-\frac{1}{2^{n(s+1)}}\\
 & =1-\frac{(s+1)}{2^{n}}+O\left(\frac{1}{2^{2n}}\right)-\frac{1}{2^{n(s+1)}}\\
 & =1-\frac{(s+1)}{2^{n}}+O\left(\frac{1}{2^{n\cdot\min\{1+s,2\}}}\right).
\end{align}
Thus, there exists $c,C>0$ such that 
\begin{align}
\log\prod_{n=1}^{\infty}\ell_{n}^{-1} & =\sum_{n=1}^{\infty}\log\ell_{n}^{-1}\\
 & \leq-\sum_{n=1}^{\infty}\log\left[1-\frac{c}{2^{n}}\right]\\
 & \leq C+\sum_{n=1}^{\infty}\frac{c}{2^{n}}+O\left(\frac{1}{2^{2n}}\right)\\
 & <\infty,
\end{align}
and consequently,
\[
d_{n}\dfn\prod_{j=n}^{\infty}\ell_{j}^{-1}\to1
\]
as $n\to\infty$. Now, (\ref{eq: approximate monotoncity}) implies
that 
\[
e_{n}\dfn d_{n}\cdot\gamma_{s}^{(n)}(\delta)
\]
is a non-increasing sequence which is bounded below by $0$, and thus
has a limit. Since $d_{n}\to1$ as $n\to\infty$, $\gamma_{s}^{(n)}(\delta)$
also has a limit. 
\end{IEEEproof}
\begin{IEEEproof}[Proof of Lemma \ref{lem: moments of ordinal}]
Due to monotonicity of $i^{s}$, standard bounds on sums using integrals
lead to 
\begin{align}
K_{s}(a,b) & \leq\int_{a+1}^{b+1}\frac{t^{s}}{b-a}\cdot\d t\\
 & =\frac{1}{s+1}\cdot\frac{\left[(b+1)^{s+1}-(a+1)^{s+1}\right]}{b-a}\label{eq: sums using integral upper bound}
\end{align}
and 
\begin{align}
K_{s}(a,b) & \geq\int_{a}^{b}\frac{t^{s}}{b-a}\cdot\d t\\
 & =\frac{1}{s+1}\cdot\frac{\left[b^{s+1}-a^{s+1}\right]}{b-a}.\label{eq: sums using integral lower bound}
\end{align}
The ratio between the upper and lower bound is 
\[
\kappa_{s}(a,b)\dfn\frac{(b+1)^{s+1}-(a+1)^{s+1}}{b^{s+1}-a^{s+1}}
\]
and clearly $\kappa_{s}(a_{n},b_{n})\to1$ given the premise of the
lemma. 
\end{IEEEproof}
\begin{IEEEproof}[Proof of (\ref{eq: entropy bound on guessing proof})]
From the law of total expectation 
\begin{equation}
G_{s}(X^{n}\mid f(Y^{n}))=\P(f(Y^{n})=0)\cdot G_{s}(X^{n}\mid f(Y^{n})=0)+\P(f(Y^{n})=1)\cdot G_{s}(X^{n}\mid f(Y^{n})=1).\label{eq: gussing moment - total expectation}
\end{equation}
Now, without loss of generality, we may assume that 
\[
\P(f(Y^{n})=0)\cdot G_{s}(X^{n}\mid f(Y^{n})=0)\geq\P(f(Y^{n})=1)\cdot G_{s}(X^{n}\mid f(Y^{n})=1)
\]
for all $n$, as otherwise one may consider $1-f$ which has the same
guessing-moments as $f$. As $G_{s}(X^{n}\mid f(Y^{n}))$ is unbounded
then $G_{s}(X^{n}\mid f(Y^{n})=0)$ is unbounded too. Let $\eta>0$
be given. Utilizing (\ref{eq: entropy lower bound for P}), let $g_{\eta}$
be such that $G_{s}(P)\geq g_{\eta}$ ensures that 
\[
G_{s}(P)\geq(1-\eta)\cdot\Psi_{s}\cdot2^{sH(P)}.
\]
Since $G_{s}(X^{n}\mid f(Y^{n})=0)$ is unbounded by our assumption,
we get from (\ref{eq: entropy lower bound for P}) that there exists
$n_{0}(\eta)$ such that
\[
G_{s}(X^{n}\mid f(Y^{n})=0)\geq(1-\eta)\cdot\Psi_{s}\cdot2^{sH(X^{n}\mid f(Y^{n})=0)}
\]
for all $n\geq n_{0}(\eta)$. Furthermore, let ${\cal N}_{+}\dfn\{G_{s}(X^{n}\mid f(Y^{n})=1)\geq g_{\eta}\}.$
If ${\cal N}_{+}$ is not empty then there exists $n_{1}'(\eta)$
such that
\[
G_{s}(X^{n}\mid f(Y^{n})=1)\geq(1-\eta)\cdot\Psi_{s}\cdot2^{sH(X^{n}\mid f(Y^{n})=1)}
\]
for all $n\geq n_{1}'(\eta)$ such that $n\in{\cal N}_{+}$. Thus,
\[
G_{s}(X^{n}\mid f(Y^{n}))\geq(1-\eta)\cdot\Psi_{s}\cdot\left[\P(f(Y^{n})=0)\cdot2^{sH(X^{n}\mid f(Y^{n})=0)}+\P(f(Y^{n})=1)\cdot2^{sH(X^{n}\mid f(Y^{n})=1)}\right]
\]
for all $n\geq\max\{n_{0}(\eta),\;n_{1}'(\eta)\}$ such that $n\in{\cal N}_{+}$.

Now, for $\mathbb{N}\backslash{\cal N}_{+}$, the guessing moment
is bounded as $G_{s}(X^{n}\mid f(Y^{n})=1)\leq g_{\eta}$. Evidently
from (\ref{eq: entropy lower bound for P}) and the sentence that
follows, if $G_{s}(X^{n}\mid f(Y^{n})=1)$ is bounded then $H(X^{n}\mid f(Y^{n})=1)$
is also bounded. Since $H(X^{n}\mid f(Y^{n})=0)$ must be unbounded,
there exists $n_{1}''(\eta)$ such that 
\[
G_{s}(X^{n}\mid f(Y^{n}))\geq(1-\eta/2)\cdot\Psi_{s}\cdot\left[\P(f(Y^{n})=0)\cdot2^{sH(X^{n}\mid f(Y^{n})=0)}+\P(f(Y^{n})=1)\cdot2^{sH(X^{n}\mid f(Y^{n})=1)}\right]
\]
for all $n\geq\max\{n_{0}(\eta),\;n_{1}''(\eta)\}$ such that $n\in\mathbb{N}\backslash{\cal N}_{+}$. 
\end{IEEEproof}
\begin{IEEEproof}[Proof of (\ref{eq: greedy dictator gussing rule})]
Let us evaluate the posterior probability conditioned on $\gdic(Y^{n})=0$.
Since $\gdic$ is balanced, Bayes law implies that 
\begin{align}
 & \P(X^{n}=x^{n}\mid\gdic(Y^{n})=0)\nonumber \\
 & =2^{-(n-1)}\cdot\P(\gdic(Y^{n})=0\mid X^{n}=x^{n})\\
 & =2^{-(n-1)}\cdot\sum_{i=1}^{n+1}\P\left(k(y^{n})=i\mid X^{n}=x^{n}\right)\cdot\P(\gdic(Y^{n})=0\mid X^{n}=x^{n},k(y^{n})=i)\\
 & =2^{-(n-1)}\cdot\left\{ \sum_{i=1}^{n}(1-\epsilon)\epsilon^{i-1}\cdot\I\left\{ x_{i}=0\right\} +\epsilon^{n}\cdot\frac{1}{2}\right\} .
\end{align}
This immediately leads to the guessing rule in (\ref{eq: greedy dictator gussing rule}).
As stated in the beginning of Section \ref{sec: General source-channel},
the guessing rule for $\gdic(Y^{n})=1$ is on reversed order. 
\end{IEEEproof}
\begin{IEEEproof}[Proof of Proposition \ref{prop: local minima of gdic}]
We denote the lexicographic order by $\ord_{\lex}$. Assume that
$\gdic(Y^{n})=0$ and that $\ord_{\lex}(x^{n})\leq\ord_{\lex}(z^{n})$.
Then, there exists $j\in[n]$ such that $x^{j-1}=z^{j-1}$ (where
$x^{0}$ is the empty string) and $x_{j}=0<z_{j}=1$. Then,
\begin{align}
 & \P(X^{n}=x^{n}\mid\gdic(Y^{n})=0)-\P(X^{n}=z^{n}\mid\gdic(Y^{n})=0)\nonumber \\
 & =\epsilon^{j-1}+\sum_{i=j+1}^{n}\epsilon^{i-1}\cdot\left(z_{i}-x_{i}\right)\\
 & \geq\epsilon^{j-1}-\sum_{i=j+1}^{n}\epsilon^{i-1}\\
 & =\frac{\epsilon^{j-1}}{1-\epsilon}\left(1-2\epsilon+\epsilon^{n-j+1}\right)\\
 & \geq0.
\end{align}
This proves the first statement of the proposition. Now, let $\ord_{0}$
($\ord_{1}$) be the guessing order given that the received bit is
$0$ (resp. 1), and let $f$ the Boolean function (which are not necessarily
optimal). Then, from (\ref{eq: posterior over BEC 0}) and (\ref{eq: posterior over BEC 1})
\begin{align}
 & G_{1}(X^{n}\mid f(Y^{n}))\nonumber \\
 & =\sum_{x^{n}}\P(X^{n}=x^{n},f(Y^{n})=0)\cdot\ord_{0}(x^{n})+\P(X^{n}=x^{n},f(Y^{n})=1)\cdot\ord_{1}(x^{n})\\
 & =2^{-n}\cdot\sum_{S\subseteq[n]}\Pr(S)\sum_{x^{n}}\left[\left(1-f_{S}(x^{n})\right)\cdot\ord_{0}(x^{n})+f_{S}(x^{n})\cdot\ord_{1}(x^{n})\right]\\
 & =2^{-n}\cdot\sum_{S\subseteq[n]}\Pr(S)\sum_{x^{S}}\left[\left(1-f_{S}(x^{n})\right)\cdot\pord_{0}(x^{S}||S)+f_{S}(x^{n})\cdot\pord_{1}(x^{S}||S)\right]\\
 & \geq2^{-n}\cdot\sum_{S\subseteq[n]}\Pr(S)\sum_{x^{n}}\min\left\{ \pord_{0}(x^{S}||S),\pord_{1}(x^{S}||S)\right\} ,\label{eq: projected order guessing inequality}
\end{align}
where for $b\in\{0,1\}$, the \emph{projected orders} are defined
as
\[
\pord_{b}(x^{S}||S)\dfn\sum_{x^{S^{c}}}\ord_{b}(x^{n}).
\]
It is easy to verify that if $\ord_{0}$ ($\ord_{1}$) is the lexicographic
(resp. revered lexicographic) order then the Greedy Dictator achieves
(\ref{eq: projected order guessing inequality}) with equality, due
to the following simple property: If $\ord_{\lex}(x^{n})<\ord_{\lex}(z^{n})$
then 
\[
\sum_{x^{S^{c}}}\ord_{\lex}(x^{n})\leq\sum_{x^{S^{c}}}\ord_{\lex}(z^{n})
\]
for all $S\in[n]$. This can be proved by induction over $n$. For
$n=1$ the claim is easily asserted. Suppose it holds for $n-1$,
and let us verify it for $n$. If $1\in S$ then whenever $\ord_{\lex}(x^{n})<\ord_{\lex}(z^{n})$
\begin{align}
\sum_{x^{S^{c}}}\ord_{\lex}(x^{n}) & =\sum_{x^{S^{c}}}\ord_{\lex}(x_{1},x_{2}^{n})\\
 & =x_{1}\cdot2^{n-1}+\sum_{x^{S^{c}}}\ord_{\lex}(x_{2}^{n})\\
 & \leq z_{1}\cdot2^{n-1}+\sum_{z^{S^{c}}}\ord_{\lex}(z_{2}^{n})\\
 & =\sum_{z^{S^{c}}}\ord_{\lex}(z^{n})
\end{align}
where the inequality follows from the induction assumption and since
$x_{1}\leq z_{1}$. If $1\notin S$ then, similarly,
\begin{align}
\sum_{x^{S^{c}}}\ord_{\lex}(x^{n}) & =\sum_{x^{S^{c}\backslash\{1\}}}\left[2^{n-1}+2\cdot\ord_{\lex}(x_{2}^{n})\right]\\
 & \leq\sum_{z^{S^{c}\backslash\{1\}}}\left[2^{n-1}+2\cdot\ord_{\lex}(z_{2}^{n})\right]\\
 & =\sum_{z^{S^{c}}}\ord_{\lex}(z^{n}).
\end{align}
\end{IEEEproof}
\begin{IEEEproof}[Proof of Theorem \ref{thm: Greedy dictator}]
We denote the lexicographic order by $\ord_{\lex}$. Then, 
\begin{align}
G_{1}(X^{n}\mid\gdic(Y^{n})) & =G_{1}(X^{n}\mid\gdic(Y^{n})=0)\\
 & \leq\sum_{x^{n}}\P(X^{n}=x^{n}\mid\gdic(Y^{n})=0)\cdot\ord_{\lex}(x^{n})\\
 & =2^{-(n-1)}\cdot\sum_{x^{n}}\sum_{i=1}^{n}(1-\epsilon)\epsilon^{i-1}\cdot\I\left\{ x_{i}=0\right\} \cdot\ord_{\lex}(x^{n})+\epsilon^{n}K_{1}(2^{n})\\
 & =2^{-(n-1)}\cdot(1-\epsilon)\sum_{i=1}^{n}\epsilon^{i-1}\cdot\sum_{x^{n}}\I\left\{ x_{i}=0\right\} \cdot\ord_{\lex}(x^{n})+\epsilon^{n}K_{1}(2^{n})\\
 & =(1-\epsilon)J_{n}+\epsilon^{n}K_{1}(2^{n}).
\end{align}
where $J_{1}\dfn1/2$ and for $n\geq2$
\begin{align}
J_{n} & \dfn2^{-(n-1)}\cdot\sum_{i=1}^{n}\epsilon^{i-1}\cdot\sum_{x^{n}}\I\left\{ x_{i}=0\right\} \cdot\ord_{\lex}(x^{n})\\
 & =2^{-(n-1)}\sum_{x^{n}}\I\left\{ x_{i}=0\right\} \cdot\ord_{\lex}(x^{n})+2^{-(n-1)}\sum_{i=2}^{n}\epsilon^{i-1}\cdot\sum_{x^{n}}\I\left\{ x_{i}=0\right\} \cdot\ord_{\lex}(x^{n})\\
 & =K_{1}(2^{n-1})\nonumber \\
 & \hphantom{=}+2^{-(n-1)}\sum_{i=2}^{n}\epsilon^{i-1}\cdot\sum_{x_{2}^{n}}\left[\I\left\{ x_{1}=0,x_{i}=0\right\} \cdot\ord_{\lex}(0,x_{2}^{n})+\I\left\{ x_{1}=1,x_{i}=0\right\} \cdot\ord_{\lex}(1,x_{2}^{n})\right]\\
 & =K_{1}(2^{n-1})+2^{-(n-1)}\epsilon\sum_{i=1}^{n-1}\epsilon^{i-1}\cdot\sum_{x^{n-1}}\I\left\{ x_{i}=0\right\} \ord_{\lex}(x^{n-1})\nonumber \\
 & \hphantom{=}+2^{-(n-1)}\epsilon\sum_{i=1}^{n-1}\epsilon^{i-1}\cdot\sum_{x^{n-1}}\I\left\{ x_{i}=0\right\} \left[2^{n-1}+\ord_{\lex}(x^{n-1})\right]\\
 & =K_{1}(2^{n-1})+\epsilon J_{n-1}+\sum_{i=1}^{n-1}\epsilon^{i}\cdot\sum_{x^{n-1}}\I\left\{ x_{i}=0\right\} \\
 & =K_{1}(2^{n-1})+\epsilon J_{n-1}+2^{n-2}\cdot\frac{\epsilon-\epsilon^{n}}{1-\epsilon}.
\end{align}
So
\begin{align}
J_{n} & =K_{1}(2^{n-1})+\epsilon\left[K_{1}(2^{n-2})+\epsilon J_{n-2}+2^{n-3}\cdot\frac{\epsilon-\epsilon^{n-1}}{1-\epsilon}\right]+2^{n-2}\cdot\frac{\epsilon-\epsilon^{n}}{1-\epsilon}\\
 & =K_{1}(2^{n-1})+\epsilon K_{1}(2^{n-2})+\epsilon^{2}J_{n-2}+2^{n-3}\cdot\frac{\epsilon^{2}-\epsilon^{n}}{1-\epsilon}+2^{n-2}\cdot\frac{\epsilon-\epsilon^{n}}{1-\epsilon}\\
 & =\sum_{i=1}^{n}\epsilon^{i-1}K_{1}(2^{n-i})+\frac{1}{1-\epsilon}\sum_{i=1}^{n}2^{i-2}\cdot(\epsilon^{n-i+1}-\epsilon^{n}).
\end{align}
Hence, 
\[
G_{1}(X^{n}\mid\gdic(Y^{n}))\leq(1-\epsilon)\sum_{i=1}^{n}\epsilon^{i-1}K_{1}(2^{n-i})+\sum_{i=1}^{n}2^{i-2}\cdot\left(\epsilon^{n-i+1}-\epsilon^{n}\right)+\epsilon^{n}K_{1}(2^{n}).
\]
Noting that $K_{1}(M)=M/2+1/2$, we get 
\begin{align}
 & G_{1}(X^{n}\mid\gdic(Y^{n}))\nonumber \\
 & \leq2^{n-1}\frac{(1-\epsilon)}{\epsilon}\sum_{i=1}^{n}\left(\frac{\epsilon}{2}\right)^{i}+\frac{(1-\epsilon)(1-\epsilon^{n})}{2\epsilon}+\frac{1}{4}\sum_{i=1}^{n}\left(\frac{2}{\epsilon}\right)^{i}\cdot\epsilon^{n+1}-\frac{1}{2}\left(2^{n}-1\right)\epsilon^{n}+2^{n-1}\epsilon^{n}+\frac{\epsilon^{n}}{2}\\
 & =\frac{1}{2-\epsilon}\left(2^{n-1}+\frac{\epsilon^{n}}{2}(1-\epsilon)\right)+\frac{(1-\epsilon)(1-\epsilon^{n})}{2\epsilon}\\
 & \cong\frac{2^{n-1}}{2-\epsilon}.
\end{align}
\end{IEEEproof}
\bibliographystyle{ieeetr}
\bibliography{Guessing_Boolean_helper}

\begin{thebibliography}{10}

\bibitem{massey1994guessing}
J.~L. Massey, ``Guessing and entropy,'' in {\em IEEE International Symposium on
  Information Theory}, p.~204, 1994.

\bibitem{arikan1996inequality}
E.~Arikan, ``An inequality on guessing and its application to sequential
  decoding,'' {\em IEEE Transactions on Information Theory}, vol.~42,
  pp.~99--105, January 1996.

\bibitem{arikan1998guessing}
E.~Arikan and N.~Merhav, ``Guessing subject to distortion,'' {\em IEEE
  Transactions on Information Theory}, vol.~44, pp.~1041--1056, March 1998.

\bibitem{merhav1999shannon}
N.~Merhav and E.~Arikan, ``The {S}hannon cipher system with a guessing
  wiretapper,'' {\em IEEE Transactions on Information Theory}, vol.~45,
  pp.~1860--1866, June 1999.

\bibitem{hayashi2008coding}
Y.~Hayashi and H.~Yamamoto, ``Coding theorems for the {S}hannon cipher system
  with a guessing wiretapper and correlated source outputs,'' {\em IEEE
  Transactions on Information Theory}, vol.~54, no.~6, pp.~2808--2817, 2008.

\bibitem{hanawal2011shannon}
M.~K. Hanawal and R.~Sundaresan, ``The {S}hannon cipher system with a guessing
  wiretapper: General sources,'' {\em IEEE Transactions on Information Theory},
  vol.~57, no.~4, pp.~2503--2516, 2011.

\bibitem{yona2016password}
Y.~Yona and S.~Diggavi, ``Password cracking: The effect of bias on the average
  guesswork of hash functions,'' 2016.
\newblock Available online: \url{http://arxiv.org/pdf/1608.02132.pdf}.

\bibitem{christiansen2015multi}
M.~M. Christiansen, K.~R. Duffy, F.~du~Pin~Calmon, and M.~M{\'e}dard,
  ``Multi-user guesswork and brute force security,'' {\em IEEE Transactions on
  Information Theory}, vol.~61, no.~12, pp.~6876--6886, 2015.

\bibitem{sundaresan2007guessing}
R.~Sundaresan, ``Guessing under source uncertainty,'' {\em IEEE Transactions on
  Information Theory}, vol.~53, pp.~269--287, January 2007.

\bibitem{boztas1997comments}
B.~Serdar, ``Comments on "{A}n inequality on guessing and its application to
  sequential decoding",'' {\em IEEE Transactions on Information Theory},
  vol.~43, pp.~2062--2063, June 1997.

\bibitem{hanawal2011guessing}
M.~K. Hanawal and R.~Sundaresan, ``Guessing revisited: A large deviations
  approach,'' {\em IEEE Transactions on Information Theory}, vol.~57,
  pp.~70--78, January 2011.

\bibitem{wyner1973theorem}
A.~Wyner, ``A theorem on the entropy of certain binary sequences and
  applications--{II},'' {\em IEEE Transactions on Information Theory}, vol.~19,
  pp.~772--777, June 1973.

\bibitem{ahlswede1975source}
R.~Ahlswede and J.~K{\"o}rner, ``Source coding with side information and a
  converse for degraded broadcast channels,'' {\em IEEE Transactions on
  Information Theory}, vol.~21, pp.~629--637, June 1975.

\bibitem{Graczyk_Lapidoth}
R.~Graczyk and A.~Lapidoth, ``Variations on the guessing problem,'' in {\em
  IEEE International Symposium on Information Theory}, June 2018.

\bibitem{gr:17:thesis}
R.~Graczyk, ``Guessing with a helper,'' Master's thesis, ETH Zurich, 2017.

\bibitem{Bool_book}
R.~O'Donnell, {\em Analysis of {B}oolean functions}.
\newblock Cambridge University Press, 2014.

\bibitem{Boolean_conjecture}
T.~A. Courtade and G.~R. Kumar, ``Which {B}oolean functions maximize mutual
  information on noisy inputs?,'' {\em IEEE Transactions on Information
  Theory}, vol.~60, pp.~4515--4525, August 2014.

\bibitem{Ordentlich_Shayevitz_Weinstein}
O.~Ordentlich, O.~Shayevitz, and O.~Weinstein, ``An improved upper bound for
  the most informative {B}oolean function conjecture,'' May 2015.
\newblock Available online: \url{http://arxiv.org/pdf/1505.05794v2.pdf}.

\bibitem{Alex}
A.~Samorodnitsky, ``On the entropy of a noisy function,'' {\em IEEE
  Transactions on Information Theory}, vol.~62, pp.~5446--5464, October 2016.

\bibitem{kindler2015remarks}
G.~Kindler, R.~O'Donnell, and D.~Witmer, ``Remarks on the most informative
  function conjecture at fixed mean,'' 2015.
\newblock Available online: \url{http://arxiv.org/pdf/1506.03167.pdf}.

\bibitem{li2018boolean}
J.~Li and M.~M{\'e}dard, ``{B}oolean functions: noise stability,
  non-interactive correlation, and mutual information,'' 2018.
\newblock Available online: \url{http://arxiv.org/pdf/1801.04462.pdf}.

\bibitem{Chandar_Tcham}
V.~Chandar and A.~Tchamkerten, ``Most informative quantization functions,''
  tech. rep., 2014.
\newblock Available online:
  \url{http://perso.telecom-paristech.fr/~tchamker/CTAT.pdf}.

\bibitem{Boolean_quadratic}
N.~Weinberger and O.~Shayevitz, ``On the optimal {B}oolean function for
  prediction under quadratic loss,'' {\em IEEE Transactions on Information
  Theory}, vol.~63, pp.~4202--4217, July 2017.

\bibitem{burin2017reducing}
A.~Burin and O.~Shayevitz, ``Reducing guesswork via an unreliable oracle,''
  March 2017.
\newblock Available online: \url{http://arxiv.org/pdf/1703.01672.pdf}.

\bibitem{Feller}
W.~Feller, {\em An Introduction to Probability Theory and Its Applications},
  vol.~2.
\newblock New York: John Wiley \& Sons, 1971.

\bibitem{Cover:2006:EIT:1146355}
T.~M. Cover and J.~A. Thomas, {\em Elements of Information Theory}.
\newblock Wiley-Interscience, 2006.

\bibitem{wainwright2008graphical}
M.~J. Wainwright and M.~I. Jordan, ``Graphical models, exponential families,
  and variational inference,'' {\em Foundations and Trends{\textregistered} in
  Machine Learning}, vol.~1, no.~1--2, pp.~1--305, 2008.

\bibitem{Boyd}
S.~P. Boyd and L.~Vandenberghe, {\em Convex Optimization}.
\newblock Cambridge university press, 2004.

\bibitem{nadarajah2005generalized}
S.~Nadarajah, ``A generalized normal distribution,'' {\em Journal of Applied
  Statistics}, vol.~32, no.~7, pp.~685--694, 2005.

\bibitem{MGL}
A.~Wyner and J.~Ziv, ``A theorem on the entropy of certain binary sequences and
  applications--{I},'' {\em IEEE Transactions on Information Theory}, vol.~19,
  pp.~769--772, November 1973.

\bibitem{erkip1998efficiency}
E.~Erkip and T.~M. Cover, ``The efficiency of investment information,'' {\em
  IEEE Transactions on information theory}, vol.~44, no.~3, pp.~1026--1040,
  1998.

\bibitem{ahlswede1976spreading}
R.~Ahlswede and P.~G{\'a}cs, ``Spreading of sets in product spaces and
  hypercontraction of the {M}arkov operator,'' {\em The annals of probability},
  pp.~925--939, 1976.

\bibitem{raginsky2016strong}
M.~Raginsky, ``Strong data processing inequalities and ${\Phi}$--{S}obolev
  inequalities for discrete channels,'' {\em IEEE Transactions on Information
  Theory}, vol.~62, no.~6, pp.~3355--3389, 2016.

\bibitem{anantharam2014hypercontractivity}
V.~Anantharam, A.~Gohari, S.~Kamath, and C.~Nair, ``On hypercontractivity and a
  data processing inequality,'' in {\em Proc. IEEE International Symposium
  Information Theory (ISIT)}, pp.~3022--3026, 2014.

\bibitem{anantharam2013maximal}
V.~Anantharam, A.~Gohari, S.~Kamath, and C.~Nair, ``On maximal correlation,
  hypercontractivity, and the data processing inequality studied by {E}rkip and
  {C}over,'' 2013.
\newblock Available online: \url{http://arxiv.org/pdf/1304.6133.pdf}.

\end{thebibliography}

\end{document}